\documentclass[11pt,a4paper]{amsart}
\usepackage[dvipsnames]{xcolor}
\usepackage[utf8]{inputenc}
\usepackage{amsfonts,amssymb,graphicx,epsfig,pdflscape,multirow,theorem,tikz,tikz-cd,braket}
\usepackage{subfigure}
\usepackage[margin=25mm]{geometry}
\usepackage{parskip}
\usepackage{bbm}
\usepackage{cite}
\usepackage{mathtools}
\usepackage[colorlinks=true,citecolor=red,linkcolor=blue]{hyperref}
\numberwithin{equation}{section}
\newcommand{\nc}{\newcommand}
\nc{\bib}{\bibitem}
\nc{\nn}{\nonumber\\ }
\nc{\chit}{\raisebox{0.25ex}{$\chi$}}
\nc{\chih}{\raisebox{0.25ex}{$\hat\chi$}}
\nc{\g}{\mathfrak{g}}
\nc{\gh}{\widehat{\mathfrak{g}}}
\nc{\Ac}{\mathcal{A}}
\nc{\Bc}{\mathcal{B}}
\nc{\Acb}{\bar{\mathcal{A}}}
\nc{\Ic}{\mathcal{I}}
\nc{\Nc}{\mathcal{N}}
\nc{\Oc}{\mathcal{O}}
\nc{\Qc}{\mathcal{Q}}
\nc{\Vc}{\mathcal{V}}
\nc{\NS}{\mathrm{NS}}
\nc{\Vir}{\mathfrak{Vir}}
\newcommand{\asl}{\widehat{\mathfrak{sl}}}
\nc{\Virb}{\overline{\mathfrak{Vir}}}
\nc{\pa}{\partial}
\nc{\eps}{\epsilon}
\nc{\Tb}{\bar{T}}
\nc{\Wb}{\bar{W}}
\nc{\cb}{\bar{c}}
\nc{\Ab}{\bar{A}}
\nc{\Bb}{\bar{B}}
\nc{\Cb}{\bar{C}}
\nc{\Jb}{\bar{J}}
\nc{\zb}{\bar{z}}
\nc{\kb}{\bar{k}}
\nc{\Gc}{\mathcal{G}}
\nc{\C}{\mathbb{C}}
\nc{\al}{\alpha}
\nc{\hh}{\widehat{\mathfrak{h}}}
\nc{\bet}{\beta}

\newcommand{\id}{\mathbbm{1}}
\newcommand{\Z}{\mathbb{Z}}

\nc{\dotNO}[1]{:\! {#1} \!:\!}
\nc{\Ah}{\widehat{A}}
\nc{\Bh}{\widehat{B}}
\nc{\Ch}{\widehat{C}}
\nc{\Jh}{\widehat{J}}
\nc{\Th}{\widehat{T}}
\nc{\Uh}{\widehat{U}}
\nc{\Vh}{\widehat{V}}
\nc{\Wh}{\widehat{W}}
\nc{\La}{\Lambda}
\nc{\Lh}{\hat{\Lambda}}
\nc{\I}{\mathbb{I}}
\nc{\ii}{\mathrm{i}}

\makeatletter
\renewcommand\author@andify{%
  \nxandlist {\unskip ,\penalty-1 \space\ignorespaces}%
    {\unskip {} \@@and~}%
    {\unskip \penalty-2 \space \@@and~}%
}
\makeatother

\begin{document}
\title{Staggered modules of $N=2$ superconformal minimal models}
\author[C~Raymond]{Christopher Raymond}
\address[Christopher Raymond]{Mathematical Sciences Institute \\
Australian National University \\
Acton, Australia, 2601.
}
\email{christopher.raymond@anu.edu.au}

\author[D~Ridout]{David Ridout}
\address[David Ridout]{
School of Mathematics and Statistics \\
University of Melbourne \\
Parkville, Australia, 3010.
}
\email{david.ridout@unimelb.edu.au}

\author[J~Rasmussen]{J\o rgen Rasmussen}
\address[J\o rgen Rasmussen]{School of Mathematics and Physics \\
The University of Queensland \\
St Lucia, Australia, 4072.
}
\email{j.rasmussen@uq.edu.au}
\date{}

\begin{abstract}
We investigate a class of reducible yet indecomposable modules over the $N=2$ superconformal algebras. These so-called staggered modules exhibit a non-diagonalisable action of the Virasoro mode $L_{0}$. Using recent results on the coset construction of $N=2$ minimal models, we explicitly construct such modules for central charges $c = -1$ and $c = -6$. We also describe spectral-flow orbits and symmetries of the families of staggered modules which arise via the coset.
\end{abstract}

\maketitle

\section{Introduction}
Logarithmic conformal field theories (CFTs) are intimately related to the representation theory of vertex operator algebras (VOAs), see \cite{CR13} for a review. In the traditional setting of rational CFT, the representations of the chiral symmetry algebra are completely reducible and are made up of finitely many irreducible summands. In logarithmic CFT, we no longer have complete reducibility; the state space includes reducible yet indecomposable modules. A subclass of these modules, known as staggered modules, has the property that the Virasoro zero-mode operator $L_{0}$ exhibits rank-two Jordan blocks \cite{R96,GK961}. The presence of these modules in the theory leads to the logarithmic divergences in correlation functions after which logarithmic CFT is named \cite{Gur93}.

While logarithmic CFTs are fundamentally non-unitary, this does not preclude them from being physically important. Many examples of logarithmic CFTs have appeared in the literature. A non-exhaustive list of examples includes theories based on the Virasoro algebra \cite{GK961,EF06,PRZ06,RS07,MR07}, fractional level WZW models based on $\asl(2)$ \cite{Gab01,Ada05,Ras06,Rid10,Ras19} and $\asl(3)$ \cite{Ada14}, WZW models based on Lie supergroups \cite{Kau00,SS05,SS06,CR13a,CR13b,CKLR18,CMY20}, bosonic ghost systems \cite{RW151,Ada19,All20}, triplet algebras \cite{GK962,GK99,Gab03,FGST06,Ras09,TW13,RW14a}, the Heisenberg algebra \cite{Cro16}, and the $N=1$ superconformal algebras \cite{PRT14,CRR15,CR15}.

Many examples of logarithmic CFTs admit a unifying understanding within the standard-module formalism of \cite{CR13,RW152}. This formalism posits that certain VOAs admit a distinguished set of representations, called standard modules, whose characters form a (topological) basis for the space of all (physically relevant) characters and moreover carry a representation of the modular group $SL(2,\Z)$. The parameter space of the standard modules is a measurable space and the modules are divided into two classes: typical ones which are irreducible and atypical ones which are not. The latter are responsible for the logarithmic properties of the correlation functions.

Determining a Verlinde formula \cite{Ver88} for Grothendieck (character) fusion in logarithmic CFTs is a key motivation of the standard-module formalism. Early attempts at applying the Verlinde formula to CFTs such as fractional-level WZW models resulted in negative integer fusion coefficients \cite{KS88,MW90}.
The logarithmic nature of these theories was not realised until much later \cite{Gab01} and it took many more years before the subtle issue that led to the observed negative integer fusion coefficients was determined \cite{Rid09} and cured \cite{CR12,CR132}.
The standard-module formalism posits a Verlinde-like formula for the (Grothendieck) fusion coefficients of the modules of the CFT. Other Verlinde-like formulae for logarithmic CFTs have been studied in \cite{FHST04,FK07,GR08,GT09,PRR10,R10,PR11}.

A categorical understanding of the representation theory of logarithmic CFTs is an active area of investigation, see \cite{CHY17} for example. The category will be non-finite and non-semisimple in general, yet admit a tensor structure with some generalised notion of modularity. The indecomposable projective objects in such a category should then be the typical standard modules and the projective covers of the atypical standard modules. It has been conjectured in many cases, see \cite{CLRW19} for example, that these atypical projective covers are staggered modules. However, this has so far only been established rigorously in a few cases \cite{All20,CMY20}.

In this paper, we use results on fractional-level $\asl(2)$ models \cite{AM95,Gab01,LMRS02,LMRS04,Rid09,Rid10a,Rid10,RW153,Ada19a,KR19}, especially those established within the standard-module formalism \cite{CR12,CR132}. This includes the classification of irreducible relaxed highest-weight modules \cite{AM95,RW153}, the identification of the standard modules \cite{CR12,CR132}, the computation of their characters \cite{KR19} and the determination of their modular transformations and Grothendieck fusion rules \cite{CR132}. We note that the latter are consistent with the genuine fusion rules that have been computed for the models with levels $k = -\frac{4}{3}$ \cite{Gab01,CR12} and $k = -\frac{1}{2}$ \cite{Rid10}. Moreover, the atypical staggered modules that arise in these models have been studied in detail.

The $\asl(2)$ minimal models and the $N=2$ superconformal minimal models are related by a well-known coset (commutant) construction \cite{KS891,KS892,FST97}. This coset realises the $N=2$ superconformal minimal-model VOA $M(u,v)$ as
\begin{equation}
    M(u,v) = \frac{A_{1}(u,v) \otimes \mathfrak{bc}}{H}, \quad u,v \in \Z, \ u \geq 2, \ v \geq 1, \ \gcd(u,v)=1,
    \label{Mcoset}
\end{equation}
where $A_{1}(u,v)$ is the admissible-level $\asl(2)$ minimal-model VOA of level $k = -2 + \frac{u}{v}$, $\mathfrak{bc}$ is the fermionic ghost VOA and $H$ is the Heisenberg VOA. In the case where $k$ is non-integer, so $v>1$, the $M(u,v)$ minimal model is non-unitary with reducible but indecomposable modules \cite{CLRW19}.

Recent work \cite{CKLR16} on general Heisenberg cosets provides a concrete dictionary between the representation theories of the $\asl(2)$ and $N=2$ minimal models, in both the unitary and non-unitary cases. In particular, this dictionary is structure-preserving --- the known (or conjectured) structures of reducible yet indecomposable modules over the $\asl(2)$ minimal-model algebras $A_{1}(u,v)$ lead to known (or conjectured) structures for the reducible yet indecomposable modules over $M(u,v)$.

This dictionary was explored in \cite{Liu19,CLRW19}, where the irreducible highest-weight modules of the $N=2$ minimal-model VOAs $M(u,v)$ were classified, their characters determined and their (Grothendieck) fusion rules computed. While this provides a lot of new knowledge about these models, especially the non-unitary ones, it leaves many questions unanswered simply because the answers are not yet known for the $\asl(2)$ minimal models.  In particular, a proper study of staggered $M(u,v)$-modules is needed to improve our knowledge of the non-unitary $N=2$ minimal models and to provide new insights into their $\asl(2)$ cousins.

In this paper, building on the results of \cite{Liu19,CLRW19,CKLR16}, we present the first investigation of staggered modules over the $N=2$ superconformal algebras. We begin in Section~\ref{Sec:N=2} with an introduction to the $N=2$ superconformal algebras, their highest-weight representation theory, and their conjugation and spectral-flow morphisms. In Section~\ref{Sec:StagMods}, we introduce the rank-$2$ staggered modules discussed conjecturally in \cite{CLRW19}. We recall their Loewy diagrams and introduce the structure parameters known as logarithmic couplings \cite{MR07,KR09} in the $N=2$ setting. Section~\ref{Sec:M(3,2)} presents explicit examples of staggered modules for the minimal model $M(3,2)$. We begin with this example, as the corresponding $\asl(2)$ minimal model $A_{1}(3,2)$ (level $k = -\frac{1}{2}$) is one of the best understood logarithmic CFTs, see \cite{LMRS04,Rid09,Rid10a,Rid10,CR12,CR13}.

In Section~\ref{Sec:GenSyms}, we discuss the effects of the conjugation and spectral-flow morphisms on general $N=2$ staggered modules. We verify that the $N=2$ staggered-module families produced by the coset branching rules are closed under the action of spectral flow and describe how spectral flow affects the logarithmic couplings. Moreover, we demonstrate a symmetry between particular parameter values that allows identification of Loewy diagrams between staggered module families.

Section~\ref{Sec:M(2,3)} gives two further explicit examples of staggered modules in the minimal model $M(2,3)$. This minimal model is related to the $\asl(2)$ minimal model $A_{1}(2,3)$ for which the staggered modules have been constructed \cite{Gab01,Ada05,CR12}.  Finally, Section~\ref{Sec:VacuumStag} addresses a family of staggered modules in which the vacuum module of $M(u,v)$ appears as a submodule (and as a quotient). Such a family exists in all non-unitary $N=2$ minimal models $M(u,v)$ and we refer to its members as the vacuum staggered modules. These modules are closely tied to the Neveu-Schwarz sector modules investigated in Sections~\ref{Sec:M(3,2)} and \ref{Sec:M(2,3)}. We determine the structures of these modules in general and note that their spectral flows provide illustrative examples of some of the general results described in Section~\ref{Sec:GenSyms}.

\section{$N=2$ superconformal algebras}
\label{Sec:N=2}
The $N=2$ superconformal algebras are a family of vertex operator superalgebras (VOSAs) parametrised by the central charge $c \in \C$. The algebras are strongly generated by four fields: $T(z), J(z)$ are bosonic, and $G^{\pm}(z)$ are fermionic. The field $T(z)$ is the Virasoro field and generates conformal symmetry. The other generating fields are Virasoro primary fields with conformal weights $1, \frac{3}{2}, \frac{3}{2}$, respectively. The defining operator product expansions (OPEs) for the universal $N=2$ VOSA at central charge $c$ are given by
\begin{equation} \label{N=2OPE}
	\begin{gathered}
    T(z)T(w) \sim \frac{\tfrac{c}{2} \id}{(z-w)^4} + \frac{2T(w)}{(z-w)^2} + \frac{ \partial T(w)}{z-w}, \quad J(z)J(w) \sim \frac{\tfrac{c}{3} \id }{(z-w)^2}, \\
    T(z)J(w) \sim \frac{J(w)}{(z-w)^2} + \frac{\partial J(w)}{z-w}, \quad T(z)G^{\pm}(w) \sim \frac{\tfrac{3}{2} G^{\pm}(w)}{(z-w)^2} + \frac{\partial G^{\pm}(w)}{z-w}, \\
    J(z)G^{\pm}(w) \sim \frac{\pm G^{\pm}(w)}{z-w}, \quad
    G^{\pm}(z)G^{\mp}(w) \sim \frac{\tfrac{2c}{3}\id}{(z-w)^3} \pm \frac{2J(w)}{(z-w)^2} + \frac{2T(w) \pm \partial J(w)}{z-w}, \\
    G^{\pm}(z) G^{\pm}(w) \sim 0,
	\end{gathered}
\end{equation}
where $\id$ is the identity field corresponding to the vacuum vector.

The universal $N=2$ VOSA is non-simple if and only if \cite{GK06}
\begin{equation}
    c = 3\left(1 - \frac{2}{t}\right), \qquad t = \frac{u}{v}, \qquad \operatorname{gcd}(u,v) = 1, \ u \in \Z_{\geq 2}, \ v \in \Z_{\geq 1}.
    \label{admissc}
\end{equation}
We are primarily interested in the case in which the universal $N=2$ VOSA is not simple. We refer to the unique simple quotient of the universal VOSA as the $N=2$ minimal-model algebra and denote it by $M(u,v)$.

The modes of the generating fields form an infinite-dimensional Lie superalgebra. Several choices of mode labelling are possible, depending on the boundary conditions chosen for the fields in the theory. In this paper, we focus on two choices: the Neveu-Schwarz algebra
\begin{equation}
    \mathrm{NS} = \operatorname{span}_{\C} \{ L_{n}, J_{m}, G^{+}_{r}, G^{-}_{s}, \id \ | \ n, m \in \Z, \ r,s \in \Z + \tfrac{1}{2} \}
\end{equation}
and the Ramond algebra
\begin{equation}
\mathrm{R} = \operatorname{span}_{\C} \{ L_{n}, J_{m}, G^{+}_{r}, G^{-}_{s}, \id \ | \ n, m \in \Z, \ r,s \in \Z \}.
\end{equation}
In this setting, the element $\id$ is a central element of the Lie superalgebra. The non-zero (anti)commutation relations defining the algebra are
\begin{equation}
	\begin{gathered}
    \left[ L_{m}, L_{n} \right] = (m-n)L_{m+n} + \frac{c}{12}m(m^2-1)\delta_{m+n,0}\id, \\
    \{ G^{\pm}_{r}, G^{\mp}_{s} \} = 2L_{r+s} \pm (r - s) J_{r+s} + \frac{c}{12}(4r^2-1)\delta_{r+s,0}\id, \\
    \left[ L_{m}, J_{n} \right] = -nJ_{m+n}, \quad \left[ J_{m}, J_{n} \right] = \frac{c}{3}m\delta_{m+n,0}\id, \\
    \left[ L_{m}, G_{r}^{\pm} \right] = \left(\frac{m}{2} - r \right)G^{\pm}_{m+r}, \quad \left[ J_{m}, G_{r}^{\pm} \right] = \pm G^{\pm}_{m+r}.
	\end{gathered}
\end{equation}

\subsection{Representation theory}
Here we introduce highest-weight modules over the $N=2$ superconformal algebras. Both the Neveu-Schwarz and Ramond algebras admit a triangular decomposition
\begin{equation}
    \mathfrak{g} = \mathfrak{g}_{-} \oplus \mathfrak{g}_{0} \oplus \mathfrak{g}_{+}.
\end{equation}
For the Neveu-Schwarz algebra, we have
\begin{equation}
	\begin{gathered}
		\mathfrak{g}_{-} = \operatorname{span}_{\C}\{ G^{-}_{n}, G^{+}_{n}, L_{n}, J_{n} \ | \ n < 0 \}, \\
		\mathfrak{g}_{+} = \operatorname{span}_{\C}\{ G^{-}_{n}, G^{+}_{n}, L_{n}, J_{n} \ | \ n > 0 \}, \\
		\mathfrak{g}_{0} = \operatorname{span}_{\C}\{ L_{0},  J_{0},  \id \};
	\end{gathered}
\end{equation}
and for the Ramond algebra,
\begin{equation}
	\begin{gathered}
		\mathfrak{g}_{-} = \operatorname{span}_{\C}\{ G^{-}_{m}, G^{+}_{n}, L_{n}, J_{n} \ | \ m \leq 0, n < 0 \}, \\
		\mathfrak{g}_{+} = \operatorname{span}_{\C}\{ G^{+}_{m}, G^{-}_{n}, L_{n}, J_{n} \ | \ m \geq 0, n > 0 \}, \\
		\mathfrak{g}_{0} = \operatorname{span}_{\C}\{ L_{0},  J_{0},  \id \}.
	\end{gathered}
\end{equation}

Note that for the Ramond algebra, we consider $G^{-}_{0}$ as a creation operator, and $G^{+}_{0}$ an annihilator.

Consider the $1$-dimensional module $\C_{j,\Delta}^{\bullet,\pm}$ over the subalgebra $\mathfrak{g}_{0} \oplus \mathfrak{g}_{+}$, labelled by $j, \Delta \in \C, \bullet \in \{ \NS, \mathrm{R} \}$. The label $\pm$ denotes the parity of the module, which is discussed in the following paragraph. The modes $L_{0}, J_{0}, \id$ act on $\C_{j,\Delta}^{\bullet,\pm}$ as multiplication by $\Delta, j, 1$, respectively, and all other modes act trivially. One can induce from $\C_{j,\Delta}^{\bullet,\pm}$ to a Verma module of weight $(j,\Delta)$ over the Neveu-Schwarz or Ramond algebra, which we denote by $V^{\bullet,\pm}_{j,\Delta}$. We denote the highest-weight vector of $V^{\bullet,\pm}_{j,\Delta}$ by $\ket{j,\Delta}^{\pm}$. The Verma module has a unique maximal proper submodule. We let $L_{j,\Delta}^{\bullet,\pm}$ denote the irreducible quotient of $V^{\bullet,\pm}_{j,\Delta}$ by its maximal proper submodule.

As modules over a superalgebra, the $N=2$ Verma modules are $\Z_{2}$-graded, meaning that
$V^{\bullet,\pm}_{j,\Delta} = V^{0} \oplus V^{1}$, where $J_n$ and $L_n$ preserve the grading while $G^{\pm}_n$ maps between $V^0$ and $V^1$. The subspaces $V^{0}$ and $V^{1}$ are referred to as the even subspace and odd subspace, respectively. The parity of the module, denoted by $\pm$, is determined by whether the highest-weight vector is in the even $(\ket{j, \Delta}^{+})$ or odd $(\ket{j, \Delta}^{-})$ subspace. There is a functor $\Pi : V^{\bullet,\pm}_{j,\Delta} \to V^{\bullet,\mp}_{j,\Delta}$ which changes the parity of the module.

A singular vector $\ket{\chit}$ is a non-zero vector which is annihilated by the action of $\mathfrak{g}_{+}$ and is a simultaneous eigenvector of $\mathfrak{g}_{0}$. If a singular vector generates a proper submodule, then we refer to it as a proper singular vector. Representations of the $N=2$ superconformal algebras may also contain subsingular vectors. A subsingular vector is a vector $\ket{w}$ in a representation $V$ which is not singular in $V$, but which becomes singular in the quotient of $V$ by a submodule.

The Verma modules of the Neveu-Schwarz and Ramond algebras can be equipped with a bilinear form given by the Shapovalov form \cite{Shap72}. The highest-weight vector $\ket{j,\Delta}^{\pm}$ is normalised relative to the highest-weight vector ${}^{\pm}\!\bra{j,\Delta}$ of the contragredient dual module $\left(V^{\pm,\bullet}_{j,\Delta} \right)^{\ast}$ such that ${}^{\pm}\!\braket{j,\Delta|j,\Delta}^{\pm} = 1$. Moreover, the adjoint is given by
\begin{equation}
    L_{n}^{\dagger} = L_{-n}, \quad J_{n}^{\dagger} = J_{-n}, \quad \big( G^{\pm}_{r}\big)^{\dagger} = G^{\mp}_{-r}, \quad \id^{\dagger} = \id.
    \label{N=2adjoint}
\end{equation}
Proper submodules are null with respect to this form.

A well-known result of \cite{BFK86} is the determinant formula for the Gram matrix of inner products for Verma modules over both the Neveu-Schwarz and Ramond algebras. For Neveu-Schwarz Verma modules, the determinant for the Gram matrix at grade $n$ is given by
\begin{equation}
    \det\nolimits_{n}(V_{j,\Delta}^{\mathrm{NS},\pm}) = \mathop{\prod_{r \in \Z_{>0}} \prod_{s \in 2\Z_{>0}}}\limits_{1\leq rs \leq 2n} \left(f^{\NS}_{r,s} \right)^{P_{\NS}(n-rs/2,m)} \cdot \prod_{q \in \Z+\frac{1}{2}}\left(g^{\NS}_{q} \right)^{\tilde{P}_{\NS}(n-|q|,m-\text{sgn}(q);q)},
\end{equation}
where the functions $f_{r,s}$ and $g_{q}$ are given by
\begin{equation}
    f^{\NS}_{r,s}(j,\Delta,c(t)) = \frac{(st-2r)^2}{4t^2} - j^2 - \frac{4\Delta}{t} - \frac{1}{t^2},
    \label{unchargedet}
\end{equation}
and
\begin{equation}
    g^{\NS}_{q}(j,\Delta,c(t)) = 2\Delta - 2qj - \frac{2}{t}\left(q^2 - \frac{1}{4} \right),
    \label{chargedet}
\end{equation}
with $c(t)$ given as in \eqref{admissc}. The functions $P_{\NS}$ and $\tilde{P}_{\NS}$ are partition functions whose precise form, which we shall not need, are given in \cite{BFK86}.
The vanishing of $f^{\NS}_{r,s}$ implies the existence of a singular vector with $(J_{0},L_{0})$-eigenvalue $(j,\Delta + \frac{rs}{2})$, referred to as an {\it uncharged} singular vector. Likewise, a vanishing $g^{\NS}_{q}$ implies the existence of a singular vector with weight $(j + \operatorname{sgn}(q), \Delta + q)$, referred to as a {\it charged} singular vector.

In the Ramond sector, the determinant formula is given by
\begin{equation}
    \det\nolimits_{n}(V_{j,\Delta}^{\mathrm{R},\pm}) = \mathop{\prod_{r \in \Z_{>0}} \prod_{s \in 2\Z_{>0}}}\limits_{1\leq rs \leq 2n} \left(f^{\mathrm{R}}_{r,s} \right)^{P_{\mathrm{R}}(n-rs/2,m)} \cdot \prod_{q \in \Z}\left(g^{\mathrm{R}}_{q} \right)^{\tilde{P}_{\mathrm{R}}(n-|q|,m-\text{sgn}(q);q)},
\end{equation}
where
\begin{equation}
    f^{\mathrm{R}}_{r,s}(j,\Delta,c(t)) = \frac{s^2}{4} -\left(j-\frac{1}{2}\right)^2 + \frac{1}{t}\left(\frac{1}{2} - 4\Delta - rs  \right) + \frac{1}{t^2}\left( r^2 - 1 \right),
\end{equation}
and
\begin{equation}
    g^{\mathrm{R}}_{q} = 2\Delta - \frac{2q^2}{t} - 2q\left(j-\frac{1}{2}\right) + \frac{1}{2t} - \frac{1}{4}.
\end{equation}
Again, the forms of the partition functions $P_{\mathrm{R}}$ and $\tilde{P}_{\mathrm{R}}$ may be found in \cite{BFK86}. The solutions to the vanishing equations imply the existence of singular vectors in the same way as for Neveu-Schwarz modules.

We remark that submodules of $N=2$ Verma modules are not necessarily Verma modules. An example of this, which is relevant to our later discussions, is the Verma module $V^{\mathrm{NS},+}_{0,0}$
with highest-weight vector $\ket{0,0}^+$. Setting $(j,\Delta) = (0,0)$, the function $g_{q}^{\NS}$ vanishes for $q = \pm \frac{1}{2}$, implying that the vectors $G^{\pm}_{-1/2}\ket{0,0}^+$ are singular. In the submodule generated by $\ket{w}=G^{+}_{-1/2}\ket{0,0}^+$, there is no corresponding $G^{+}_{-1/2}\ket{w}$, as $\big(G^{+}_{-1/2}\big)^2 = 0$ in the universal enveloping algebra. Thus, this submodule is not isomorphic to the Verma module $V^{\NS,-}_{1,1/2}$. Similar arguments hold for $G^{-}_{-1/2}\ket{0,0}^+$.

\subsection{Automorphisms}
The $N=2$ superconformal algebras admit a number of well-studied automorphisms \cite{SS87}. Here, we are interested in the conjugation automorphism denoted by $\gamma$, and the spectral-flow family of automorphisms denoted by $\sigma^{\ell}$, for $\ell \in \Z$.

The conjugation automorphism acts on the modes of the generating fields as
\begin{gather}
\gamma(L_{n}) = L_{n}, \quad \gamma(J_{n}) = -J_{n}, \quad \gamma(G^{\pm}_{r}) = G^{\mp}_{r}, \quad \gamma(\id) = \id,
\label{conjugationmodeaction}
\end{gather}
while the action of spectral flow is given by
\begin{gather}
    \sigma^{\ell}(L_{n}) = L_{n} - \ell J_{n} + \frac{1}{6}\ell^2 \delta_{n,0}c \id, \quad \sigma^{\ell}(J_{n}) = J_{n}+\frac{\ell}{3} \delta_{n,0}c \id, \quad \sigma^{\ell}(G^{\pm}_{r}) = G^{\pm}_{r\mp \ell}, \quad \sigma^{\ell}(\id) = \id.
    \label{spectralflowmodeaction}
\end{gather}
Spectral flow can be extended to $\ell \in \frac{1}{2}\Z$. Non-integral parameters then define a family of isomorphisms between the Neveu-Schwarz and Ramond algebras.

We remark that spectral flow is not an automorphism of the corresponding VOSA (rather, only the vertex superalgebra), as it does not preserve the conformal vector $L_{-2}\ket{0,0}^+$, where $\ket{0,0}^+$ is the vacuum vector of the vertex algebra.

Modules over an algebra may be twisted by the action of an automorphism and are then referred to as twisted modules. Suppose we have a vector-space isomorphism $\xi$ from a given $\mathfrak{g}$-module $M$ to a new vector space $\xi(M)$. This becomes a twisted module upon endowing it with the action of $\mathfrak{g}$ given by
\begin{equation} \label{gentwisteq}
    x \xi\big(\ket{v}\big) = \xi\big(\omega^{-1}(x)\ket{v}\big), \quad \ \mathrm{for} \ x \in \mathfrak{g}, \ \ket{v} \in M,
\end{equation}
where $\omega$ is an automorphism of $\mathfrak{g}$. As $\xi$ merely serves to distinguish the module $M$ from its twist, which will not be isomorphic to $M$ in general, we may replace $\xi$ by the algebra automorphism $\omega$. Moreover, we will refer to $\gamma(M)$ and $\sigma^{\ell}(M)$ as the conjugation of $M$ and the spectral flow of $M$, respectively. Thus, the twisting action of the spectral-flow automorphisms will be written as
\begin{equation}
    x \sigma^{\ell}\big(\ket{v}\big) = \sigma^{\ell} \big( \sigma^{-\ell}(x) \ket{v}\! \big), \quad x \in \mathfrak{g}.
    \label{twisteq}
\end{equation}

Combining \eqref{gentwisteq} with \eqref{conjugationmodeaction} and \eqref{spectralflowmodeaction}, we can identify the action of these automorphisms on the irreducible highest-weight modules, as in \cite{CLRW19}. For conjugate modules, we have
\begin{equation}
    \gamma\left(L_{j,\Delta}^{\NS,\pm}\right) \cong L_{-j;\Delta}^{\NS,\pm}, \qquad \gamma\left( L_{j;\Delta}^{\mathrm{R},\pm} \right) \cong \begin{cases}
    L_{-j;\Delta}^{\mathrm{R},\pm}, \qquad &\operatorname{if} \ \Delta = \frac{c}{24}, \\[10pt]
    L_{-j+1,\Delta}^{\mathrm{R},\mp}, \qquad &\mathrm{otherwise},
    \end{cases}
    \label{conjiso}
\end{equation}
while for spectral flows of modules, we have
\begin{align}
    \sigma^{\frac{1}{2}}\left( L_{j,\Delta}^{\NS,\pm} \right) \cong L_{j + \frac{c}{6},\Delta + \frac{j}{2} + \frac{c}{24}}^{\mathrm{R},\pm}, \qquad
    \sigma^{\frac{1}{2}}\left( L_{j,\Delta}^{\mathrm{R},\pm} \right) \cong \begin{cases} L_{j + \frac{c}{6},\frac{j}{2} + \frac{c}{12}}^{\NS,\pm}, \quad &\mathrm{if} \ \Delta = \frac{c}{24}, \\[10pt]
    L_{j-1+\frac{c}{6}, \Delta + \frac{j-1}{2} + \frac{c}{24}}^{\NS,\mp}, \quad &\mathrm{otherwise}.
    \end{cases}
    \label{specflowaction}
\end{align}

For Ramond sector modules, the case $\Delta = \frac{c}{24}$ is distinguished, as the vector $G^{-}_{0}\ket{j,\frac{c}{24}} \in V^{\mathrm{R},\pm}_{j,\frac{c}{24}}$ is singular. This is easily verified by computing
\begin{equation}
    G^{+}_{0}G^{-}_{0}\ket{j,\tfrac{c}{24}} = \left( 2L_{0} - \frac{c}{12}\right) \ket{j,\tfrac{c}{24}}.
\end{equation}

\section{Coset construction and staggered modules}
\label{Sec:StagMods}
We are interested in staggered modules over the $N=2$ minimal-model algebras $M(u,v)$. Following \cite{KR09,CR13}, we define a staggered module $P$ over $M(u,v)$ to be an indecomposable module described by a non-split short exact sequence
\begin{equation}
    0 \to M^{L} \to P \to M^R \to 0,
\end{equation}
where $M^L$ and $M^R$ are highest-weight $M(u,v)$-modules, and the action of $L_{0}$ on $P$ exhibits Jordan blocks of rank $2$ but not higher. The staggered modules we investigate arise through the coset construction \eqref{Mcoset} of the $N=2$ minimal models as counterparts to staggered modules over the $\asl(2)$ minimal models $A_{1}(u,v)$.

\subsection{Coset ingredients}

To better understand the background and notation, we begin this section by giving a brief introduction to the coset construction of the non-unitary $N=2$ minimal models, as described in \cite{CLRW19}. Our review will not be comprehensive, focusing only on those results which are integral to the presentation of our results. We give a brief introduction to the VOSAs which appear in the particular coset, as well as the relevant representation theories. We mention that the authors of \cite{CLRW19} analysed both the unitary and non-unitary minimal models of the $N=2$ superconformal algebras constructed via the coset \eqref{Mcoset}.

The Heisenberg VOA $H$ is generated by a single field $a(z)$ with defining OPE
\begin{equation}
    a(z)a(w) \sim \frac{2t \id}{(z-w)^2},
\end{equation}
where $t \in \C^\times$ is identified with the constant $t$ introduced in \eqref{admissc}. The commutation relations between modes of the generating field are
\begin{equation}
    \left[a_{m}, a_{n} \right] = 2t m \delta_{m+n,0} \id, \qquad m,n \in \Z.
\end{equation}
Using this particular normalisation, the energy-momentum tensor of the theory is given by
\begin{equation}
    T^{H}(z) = \frac{1}{4t}:\!aa\!:\!(z).
\end{equation}
This field generates a Virasoro algebra with central charge $1$, and $a(z)$ is a primary field of conformal weight $1$.

The highest-weight modules of $H$ are the Fock spaces of charge $p \in \C$, denoted by $F_{p}$,
with highest-weight vector $\ket{p}$. The action of the zero-modes on $\ket{p}$ is given by
\begin{equation}
    a_{0}\ket{p} = p \ket{p}, \qquad \id \ket{p} = \ket{p}, \qquad a_{n} \ket{p} = 0, \ \forall n \geq 1.
\end{equation}
The Fock spaces are irreducible for all $p$. The module $F_{0}$ is the vacuum module.

The $\mathfrak{bc}$-ghost system is a VOSA generated by the two fermionic fields $b(z)$ and $c(z)$ with defining OPEs
\begin{equation}
    b(z)c(w) \sim \frac{\id}{z-w}, \qquad b(z)b(w) \sim c(z)c(w) \sim 0.
\end{equation}
The anti-commutation relations between modes of the generating fields are
\begin{equation}
    \{ b_{m},c_{n} \} = \delta_{m+n,0}\id, \qquad \{ b_{m},b_{n} \} = \{ c_{m},c_{n} \} = 0,
\end{equation}
where $m,n \in \Z + \frac{1}{2}$ in the Neveu-Schwarz sector, and $m,n \in \Z$ in the Ramond sector. The corresponding energy-momentum tensor is
\begin{equation}
    T^{\mathfrak{bc}}(z) = \tfrac{1}{2}\big(:\!\partial b c\!:\!(z) - :\!b\partial c\!:\!(z) \big).
\end{equation}
It generates a Virasoro algebra with central charge $1$, and the fields $b(z)$ and $c(z)$ are primary fields of conformal weight $\frac{1}{2}$. The Heisenberg field $Q(z) = \, :\!bc\!:\!(z)$ acts as a charge operator with OPEs
\begin{equation}
    Q(z)b(w) \sim \frac{b(w)}{z-w}, \qquad Q(z)c(w) \sim - \frac{c(w)}{z-w}.
\end{equation}

Up to isomorphism, there are four highest-weight modules of $\mathfrak{bc}$, denoted by $N_{i}$, $i \in \{ 0,1,2,3 \}$, all of which are irreducible. The vacuum module is $N_{0}$, a Neveu-Schwarz sector module whose highest-weight vector has $Q_{0}$-eigenvalue $0$ and conformal dimension $0$. The highest-weight vector of the Ramond Verma module $N_{1}$ has $Q_{0}$-eigenvalue $\frac{1}{2}$ and conformal dimension $\frac{1}{8}$. The remaining modules are obtained by applying parity reversal: $\Pi (N_{0}) = N_{2}$ and $\Pi (N_{1}) = N_{3}$.

The VOAs associated to the affine Lie algebra $\asl(2)$ are generated by the bosonic fields $e(z)$, $h(z)$,  and $f(z)$, and have defining OPEs
\begin{gather}
    h(z)e(w) \sim \frac{2e(w)}{z-w}, \qquad h(z)h(w) \sim \frac{2k\id}{(z-w)^2}, \qquad h(z)f(w) \sim -\frac{2f(w)}{z-w}, \nn
    e(z)f(w) \sim \frac{k\id}{(z-w)^2} + \frac{h(w)}{z-w}, \qquad e(z)e(w) \sim f(z)f(w) \sim 0,
\end{gather}
where $k \in \C\setminus\{ -2 \}$ is the level of the VOA. The resulting commutation relations between modes of the generating fields are given by
\begin{gather}
    \left[ h_{m}, e_{n} \right] = 2e_{m+n}, \qquad  \left[ h_{m}, h_{n} \right] = 2km\delta_{m+n,0} \id, \qquad \left[ h_{m}, f_{n} \right] = -2f_{m+n}, \nn
    \left[ e_{m}, f_{n} \right] = h_{m+n} + km\delta_{m+n,0}\id, \qquad \left[ e_{m}, e_{n} \right] = \left[ f_{m}, f_{n} \right] = 0.
\end{gather}
The energy-momentum tensor is given by
\begin{equation}
    T^{\asl(2)}(z) = \frac{1}{2t}\left(\frac{1}{2}\!:\!hh\!:\!(z)\, + :\!ef\!:\!(z) \, + :\!fe\!:\!(z) \!\right),
\end{equation}
where $t = k + 2$ is the same $t$ as introduced for the Heisenberg algebra. We denote the modes of the field $T^{\asl(2)}(z)$ by $L^{\asl(2)}_{m}$ with $m \in \Z$. The field $T^{\asl(2)}$ generates a Virasoro algebra with central charge
\begin{equation}
    c = 3\left(1 - \frac{2}{t} \right).
\end{equation}
The universal VOA associated to $\asl(2)$ is non-simple if and only if $t = \frac{u}{v}$ for $\operatorname{gcd}(u,v) = 1, \ u \in \Z_{\geq 2}$, and $v \in \Z_{\geq 1}$ \cite{KK79,GK06}. The corresponding simple quotient for these values of $t$ is known as an $\asl(2)$ minimal model; we denote it by $A_{1}(u,v)$.

If $v=1$, the level $k$ is a non-negative integer and the corresponding minimal model is unitary. For $v > 1$, the resulting models are non-unitary. Here, we focus on the irreducible highest-weight modules over the non-unitary minimal-model algebras $A_{1}(u,v)$, as it is these that are glued together to form staggered modules. A complete classification of positive-energy $A_{1}(u,v)$-modules includes the so-called relaxed modules \cite{FST97,FSST98,RW153} which were classified in \cite{AM95}; however, we will not need these modules here.

The irreducible positive-energy representations are labelled by the $h_{0}$- and $L_{0}^{\asl(2)}$-eigenvalues, denoted by $\lambda_{r,s}$ and $\Delta_{r,s}^{\mathrm{aff}}$, respectively, of their highest-weight vectors. The classification of these representations was given in \cite{AM95}. We use the following standard parametrisations for $\lambda_{r,s}$ and $\Delta_{r,s}^{\mathrm{aff}}$:
\begin{equation}
    \lambda_{r,s} = r-1-st, \qquad \Delta_{r,s}^{\mathrm{aff}} = \frac{(r - st)^2-1}{4t}, \qquad r,s \in \Z.
\end{equation}
In this paper, we focus on the following classes of modules:
\begin{itemize}
    \item The irreducible highest-weight modules, denoted by $L_{r,0}$, where $1 \leq r \leq u-1$, with highest weights $(\lambda_{r,0}, \ \Delta_{r,0}^{\mathrm{aff}})$. As $\lambda_{r,0} \in \Z_{\geq 0}$ for all $r \geq 1$, the space of lowest-energy (minimal $L_{0}^{\asl(2)}-$eigenvalue) states forms a finite-dimensional representation over the $\mathfrak{sl}(2)$ subalgebra spanned by the modes $\{ h_{0}, e_{0}, f_{0} \}$. The vacuum module of the VOA is $L_{1,0}$.
    \item The irreducible highest-weight modules $D^{+}_{r,s}$, where $1 \leq r \leq u-1$ and $1 \leq s \leq v-1$, with highest weights $(\lambda_{r,s}, \ \Delta_{r,s}^{\mathrm{aff}})$. In this case, $\lambda_{r,s} \notin \Z$ and the space of lowest-energy states forms an infinite-dimensional irreducible Verma module over the $\mathfrak{sl}(2)$ subalgebra.
    \item The irreducible modules $D^{-}_{r,s}$, where $1 \leq r \leq u-1$ and $1 \leq s \leq v-1$. These modules are not highest-weight modules; rather, they are conjugate to the modules $D^{+}_{r,s}$. Correspondingly, the space of lowest-energy states forms an infinite-dimensional lowest-weight representation over the $\mathfrak{sl}(2)$ subalgebra, with lowest weight $-\lambda_{r,s}$.
\end{itemize}

\subsection{Coset decompositions}

For $L_{r,0} \otimes N_{i}$ and $D^{\pm}_{r,s} \otimes N_{i}$, representations over $A_{1}(u,v) \otimes \mathfrak{bc}$, we have the following branching rules for the restriction to the subalgebra $H \otimes M(u,v)$ \cite{CLRW19}:
\begin{equation}
   \left( L_{r,0} \otimes N_{i} \right) \! \downarrow \,\cong \hspace{-3ex} \bigoplus_{p \in \lambda_{r,0} + i + 2\Z} \hspace{-3ex} F_{p} \otimes {}^{[i]}C_{p;r,0}, \qquad \left(  D_{r,s}^{\pm} \otimes N_{i} \right) \! \downarrow \,\cong \hspace{-3ex} \bigoplus_{p \in \lambda_{r,s} + i + 2\Z} \hspace{-3ex} F_{p} \otimes {}^{[i]}C_{p;r,s},
    \label{branchingrule}
\end{equation}
where we recall that $i \in \{ 0,1,2,3 \}$, $F_{p}$ is the Fock space of weight $p$, and ${}^{[i]}C_{p;r,s}$ is an irreducible representation of $M(u,v)$. The irreducibility of the representation ${}^{[i]}C_{p;r,s}$ follows from the results of \cite{CKLR16}.

Through the coset, each irreducible $A_{1}(u,v)$-module gives rise to an infinite family of irreducible $M(u,v)$-modules, labelled by the weight $p$ of its partner Heisenberg Fock space.

The authors of \cite{CLRW19} give a dictionary for translating between irreducible $M(u,v)$-modules in the notation ${}^{[i]}C_{p;r,s}$ and those introduced in Section~\ref{Sec:N=2}, denoted by $L^{\bullet, \pm}_{j,\Delta}$. For the irreducible $L$-type $A_{1}(u,v)$-modules (those for which $s=0$), the corresponding modules which appear in the branching rule are given by
\small
\begin{align}
    &{}^{[0]}C_{p;r,0} \cong L_{j,\Delta}^{\NS,\bullet}, & \hspace{-2ex} &p \in \lambda_{r,0} + 2\Z, \ &
    &\left\lbrace \begin{aligned}
    &\bullet = -, &  j &= \tfrac{p}{t}+1, &  \Delta &= \Delta^{N=2}_{p;r,0} - \tfrac{p+r}{2}, &  &p\leq -r-1, \nn
    &\bullet = +, &  j &= \tfrac{p}{t}, &  \Delta &= \Delta^{N=2}_{p;r,0}, &  &1-r \leq p\leq r-1, \nn
    &\bullet = -, &  j &= \tfrac{p}{t}-1, &  \Delta &= \Delta^{N=2}_{p;r,0} + \tfrac{p-r}{2}, &  &p\geq r+1, \nn
    \end{aligned} \right. \nn
    &{}^{[1]}C_{p;r,0} \cong L_{j,\Delta}^{\mathrm{R},\bullet}, & &p \in \lambda_{r,0} + 1 + 2\Z, & \
    &\left\lbrace
    \begin{aligned}
    &\bullet = +, & \! j &= \tfrac{p}{t}+\tfrac{3}{2}, & \! \Delta &= \Delta^{N=2}_{p;r,0} - \tfrac{p+r}{2} + \tfrac{1}{8}, & &p\leq -r-2, \nn
    &\bullet = -, & \! j &= \tfrac{p}{t} + \tfrac{1}{2}, & \! \Delta &= \Delta^{N=2}_{p;r,0} + \tfrac{1}{8}, & &-r \leq p\leq r-2, \nn
    &\bullet = +, & \! j &= \tfrac{p}{t}-\tfrac{1}{2}, & \! \Delta &= \Delta^{N=2}_{p;r,0} + \tfrac{p-r}{2} + \tfrac{1}{8}, & &p\geq r, \nn
    \end{aligned}
    \right.
\end{align}
\normalsize
where we have made use of
\begin{equation}
    \Delta^{N=2}_{p;r,s} = \frac{(r-st)^2-1}{4t} - \frac{p^2}{4t}.
\end{equation}
For $1 \leq s \leq v-1$ ($D^+$-type modules), the dictionary is given by
\small
\begin{align}
    &{}^{[0]}C_{p;r,s} \cong L_{j,\Delta}^{\NS,\bullet}, &  &p \in \lambda_{r,s} + 2\Z, & \
    &\left\lbrace
    \begin{aligned}
    &\bullet = +, & j &= \tfrac{p}{t}, & \Delta &= \Delta^{N=2}_{p;r,s}, & &p\leq \lambda_{r,s}, \nn
    &\bullet = -, & j &= \tfrac{p}{t}-1, & \Delta &= \Delta^{N=2}_{p;r,s}+ \tfrac{p-\lambda_{r,s}-1}{2}, & &p \geq \lambda_{r,s}+2, \nn
    \end{aligned}
    \right. \nn
    &{}^{[1]}C_{p;r,s} \cong L_{j,\Delta}^{\mathrm{R},\bullet}, & &p \in \lambda_{r,s} + 1 + 2\Z, &
    &\left\lbrace
    \begin{aligned}
    &\bullet = -, & j &= \tfrac{p}{t} + \tfrac{1}{2}, & \Delta &= \Delta^{N=2}_{p;r,s} + \tfrac{1}{8}, & &p \leq \lambda_{r,s} - 1, \nn
    &\bullet = +, & j &= \tfrac{p}{t}-\tfrac{1}{2}, & \Delta &= \Delta^{N=2}_{p;r,s} + \tfrac{p-\lambda_{r,s} -1}{2} + \tfrac{1}{8}, & &p\geq \lambda_{r,s}+1. \nn
    \end{aligned} \right.
\end{align}
\normalsize
As the modules with $i = 2,3$ are parity reversals of those with $i=0,1$, respectively, their dictionaries follow from those given above.

For the $\asl(2)$ minimal models $A_{1}(u,v)$, staggered modules have been explicitly constructed as fusion products for $(u,v) = (3,2)$ and $(2,3)$. They are nevertheless conjectured \cite{CKLR18} to appear in all non-unitary minimal models. We remark that these conjectured minimal-model staggered modules are not the only examples of staggered modules over $\asl(2)$, see \cite{Ras19}. Under the coset, each $A_{1}(u,v)$ staggered module gives rise to a family of $M(u,v)$ staggered modules, denoted by ${}^{[i]}P_{p;r,s}$. The $M(u,v)$ staggered modules are labelled by $p$, the weight of the accompanying Fock space, $i$ the label of the $\mathfrak{bc}$-ghost system irreducible module, and $r,s$ where $1 \leq r \leq u-1$ and $0 \leq s \leq v-1$.

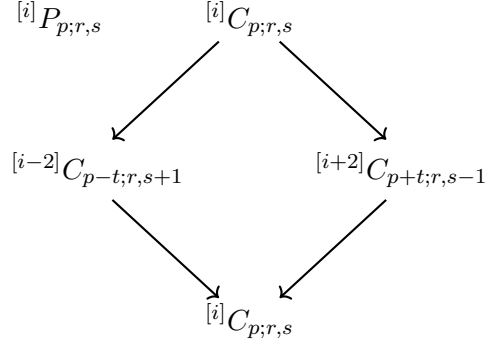
\begin{figure}
\begin{center}
\begin{tikzpicture}
\node at (-2.5,2) {${}^{[i]}P_{p;r,s}$};
\node at (-2,0) {${}^{[i-2]}C_{p-t;r,s+1}$};
\node at (0,2) {${}^{[i]}C_{p;r,s}$};
\node at (2,0) {${}^{[i+2]}C_{p+t;r,s-1}$};
\node at (0,-2) {${}^{[i]}C_{p;r,s}$};
\draw [thick,->] (0.4,1.7)--(1.8,0.4);
\draw [thick,->] (-0.4,1.7)--(-1.8,0.4);
\draw [thick,<-] (-0.4,-1.7)--(-1.8,-0.4);
\draw [thick,<-] (0.4,-1.7)--(1.8,-0.4);
\end{tikzpicture}
\caption{\textit{The Loewy diagram for a general staggered module ${}^{[i]}P_{p;r,s}$. The component modules related by arrows are irreducible $M(u,v)$-modules. All arrows are one way and indicate that the action of the algebra can map from states in one component module to states in another, but not back. We will refer to the algebra action which maps between component modules as gluing. We refer to the irreducible submodule ${}^{[i]}C_{p;r,s}$ as the socle. The component module ${}^{[i]}C_{p;r,s}$ from which all arrows point outward is referred to as the head. The remaining component modules, namely ${}^{[i-2]}C_{p-t;r,s+1}$ and ${}^{[i+2]}C_{p+t;r,s-1}$, are referred to as the left and right modules according to the $J_{0}$-weight of their component highest-weight vectors relative to that of the socle.}}
\label{Fig:GenN=2Stag}
\end{center}
\end{figure}

The Loewy diagram of ${}^{[i]}P_{p;r,s}$ displays the irreducible component modules of ${}^{[i]}P_{p;r,s}$, with arrows describing the action of the algebra which ``glues" component modules together. From the conjectured Loewy diagrams of the staggered $A_1(u,v)$-modules \cite{CKLR18}, we easily obtain those of the ${}^{[i]}P_{p;r,s}$, using the general theory of \cite{CKLR16}. These diagrams are presented in Figure~\ref{Fig:GenN=2Stag}.

In order for the module dictionary to be compatible with these Loewy diagrams for all possible $r,s,i,p$, we may be required to make replacements utilising the isomorphisms (see \cite{CLRW19})
\begin{equation}
    {}^{[i]}C_{p;r,-1}\cong{}^{[i+2]}C_{p+t;u-r,v-2},  \qquad {}^{[i]}C_{p;r,v}\cong{}^{[i-2]}C_{p-t;u-r,1}.
    \label{modsyms}
\end{equation}
The results of \cite{CKLR16} imply that the spectral flow and conjugation of an $M(u,v)$ staggered module is determined by twisting the irreducible component modules by the automorphism, maintaining the gluing structure of the Loewy diagram. Using this fact, the authors of \cite{CLRW19} deduce the following isomorphisms between irreducible $M(u,v)$-modules:
\begin{equation}
   \gamma\left( {}^{[i]}C_{p;r,s}\right) \cong {}^{[-i]}C_{-p;r,s},  \qquad \sigma^{\ell}\left({}^{[i]}C_{p;r,s} \right) \cong {}^{[i-2\ell]}C_{p-2\ell;r,s}.
    \label{moduleauts}
\end{equation}
Here, the argument of $[ \ \cdot \ ]$ is taken modulo $4$ and the parity reversal functor acts as $\Pi\left( {}^{[i]}C_{p;r,s}\right) = {}^{[i+2]}C_{p;r,s}$.

Following Figure~\ref{Fig:GenN=2Stag}, we denote the images of the highest-weight vectors of component modules in the staggered module by $\ket{v_{h}}, \ket{v_{\ell}}, \ket{v_{r}}, \ket{v_{s}}$ for the head, left, right, and socle modules, respectively. We remark that these vectors are not necessarily highest-weight vectors in the staggered module. Our diagrammatic convention is that $J_{0}$ eigenvalues increase to the right, and $L_{0}$ eigenvalues increase down the page.

\subsection{Logarithmic couplings}

The structure of a staggered module may not be completely described by its Loewy diagram. Accordingly, we require information about the gluing of the head to the left and right modules. This is described by parameters known as {\it logarithmic couplings}. The value of these parameters for a given staggered module, along with its Loewy diagram, is believed to be sufficient to characterise the module up to isomorphism. (This ``complete invariant'' property has only been proven rigorously for Virasoro staggered modules \cite{KR09}.) We motivate and define the logarithmic couplings in the remainder of this section.

The non-diagonalisability of the Virasoro zero mode $L_{0}$ on staggered $M(u,v)$-modules is captured by the relation
\begin{equation}
    L_{0}\ket{v_{h}} = \Delta_{h}\ket{v_{h}} + \ket{v_{s}},
\end{equation}
exhibiting the vector $\ket{v_{h}}$ as the logarithmic partner of $\ket{v_{s}}$. This relation does not define $\ket{v_{h}}$ in the staggered module uniquely.
There is a ``gauge freedom" whereby we are free to add any weight vector in the weight space of $\ket{v_{s}}$ to $\ket{v_{h}}$, without changing the above relation.

We are able to introduce a Shapovalov-like bilinear form on the left and right modules, normalised such that $\braket{v_{\ell}|v_{\ell}} = \braket{v_{r}|v_{r}} = 1$, and with adjoint given as in \eqref{N=2adjoint}. This form can be extended to the indecomposable module formed by including the socle as a submodule of the left and right modules. The form is therefore zero on the socle, as a proper submodule. However, as $\ket{v_{h}}$ is only defined up to multiples of $\ket{v_{s}}$ and other vectors in that weight space, we cannot normalise the form on the head. However, as we shall see below, it is possible to partially extend this form so that one of the vectors is associated to the head.

Let $U(\g)$ denote the universal enveloping algebra of the Neveu-Schwarz or Ramond algebra. The vector $\ket{v_{s}}$ satisfies
\begin{equation}
    U_{\ell}\ket{v_{\ell}} = \ket{v_{s}}, \qquad U_{r}\ket{v_{r}} = \ket{v_{s}},
    \label{soclerels}
\end{equation}
for some $U_{\ell}, U_{r} \in U(\mathfrak{g})$, such that $U_{\ell}\ket{j_{\ell}, \Delta_{\ell}} = U_{r}\ket{j_{r}, \Delta_{r}} = 0$ in the corresponding irreducible component modules.

The logarithmic couplings are then defined by the relations
\begin{equation}
    U_{\ell}^{\dagger}\ket{v_{h}} = \beta_{\ell}\ket{v_{\ell}}, \qquad U_{r}^{\dagger}\ket{v_{h}} = \beta_{r}\ket{v_{r}}, \qquad \beta_{\ell}, \beta_{r} \in \C,
    \label{betarels}
\end{equation}
with adjoint defined as in \eqref{N=2adjoint}. Hence, the logarithmic couplings can be understood as gluing parameters between the head module and the left/right modules. Alternatively, in terms of the introduced form,
\begin{equation}
\braket{v_{s}|v_{h}} = \bra{v_{\ell}}U_{\ell}^{\dagger}\ket{v_{h}} = \beta_{\ell}, \quad \braket{v_{s}|v_{h}} = \bra{v_{r}} U_{r}^{\dagger}\ket{v_{h}} = \beta_{r}.
\end{equation}
The motivation for defining the logarithmic couplings in this way is that they are invariant under the ``gauge" transformations discussed earlier. As an example to demonstrate this fact, consider the simple gauge transformation $\ket{v_{h}} \mapsto \ket{v_{h}} + \alpha \ket{v_{s}}$ for some $\alpha \in \C$. Then we have that
\begin{equation}
    \bra{v_{r}} U_{r}^{\dagger}\left( \ket{v_{h}} + \alpha \ket{v_{s}}\right) =\bra{v_{r}} U_{r}^{\dagger}\ket{v_{h}} + \alpha \bra{v_{r}} U_{r}^{\dagger}\ket{v_{s}} = \beta_{r} + 0,
\end{equation}
and a similar result follows for $\beta_{\ell}$.

To determine the numerical values of the logarithmic couplings (see also \cite{VJS11}), one uses constraints coming from vanishing singular vectors in the head module, which lift to relations between vectors in the staggered module \cite{MR08}. These constraints allow one to determine the action of the algebra on the staggered module. The logarithmic couplings, along with the Loewy diagram, are then expected to completely determine the structure of the module. The above prescription has been understood in the most detail for staggered modules over the Virasoro algebra, where it was shown in \cite{KR09} that the logarithmic couplings in fact parametrise the space of isomorphism classes of staggered modules. There is no similar result for the $N=2$ algebras as yet, but the use of logarithmic couplings is standard in logarithmic CFT.

We remark that the choice of $U_{\ell}, U_{r}$ is only fixed up to normalisation, so the logarithmic couplings depend (in a trivial way) on this choice of normalisation. As such, when introducing a module we specify our choice of normalisation for $U_{\ell}\ket{v_{\ell}} = \ket{v_{s}} = U_{r}\ket{v_{r}}$, and hence $U_{\ell}^{\dagger}, U_{r}^{\dagger}$.
\section{Staggered modules over $M(3,2)$}
\label{Sec:M(3,2)}
Here, we begin the determination and analysis of concrete examples of staggered modules over the $N=2$ superconformal algebras. We start with examples over the minimal-model VOA $M(3,2)$, for which $c = -1$. The branching rules produce families of staggered modules ${}^{[i]}P_{p;r,s}$, labelled by $p = \lambda_{r,s} + i + 2\Z$, for $(r,s) = (1,0), (1,1), (2,0), (2,1)$ and $i \in \{ 0,1,2,3 \}$. We will look at two examples in detail: ${}^{[0]}P_{0;1,0}$ from the Neveu-Schwarz sector, and ${}^{[1]}P_{\frac{3}{2};1,1}$ from the Ramond sector. These staggered modules are conjectured to be the projective covers of the irreducible highest-weight modules $L_{0,0}^{\mathrm{NS},+}$ and $L_{\frac{1}{2},\frac{5}{8}}^{\mathrm{R},+}$, respectively.

\subsection{The module ${}^{[0]}P_{0;1,0}$}
We begin by defining the gluing action of the algebra. We start with the fermionic vector $\ket{v_r}$ of weight $\left(1,-\tfrac{1}{2}\right)$ and normalise it so that $\braket{v_r | v_r } = 1$.
The vector $G^{-}_{-\frac{1}{2}} \ket{1,-\tfrac{1}{2}}^-$ is singular in the Verma module $V^{\mathrm{NS},-}_{1,-\frac{1}{2}}$ and $G^{-}_{-\frac{1}{2}}\ket{v_{r}}$ generates the socle in ${}^{[0]}P_{0;1,0}$. We may thus choose the vector $\ket{v_{s}} \in {}^{[0]}P_{0;1,0}$ to be $\ket{v_s}=G^{-}_{-\frac{1}{2}}\ket{v_{r}}$.

Similarly, we introduce a fermionic vector $\ket{v_\ell}$, with weight $\left(-1,-\tfrac{1}{2}\right)$, such that $G^{+}_{-\frac{1}{2}}\ket{v_{\ell}} = \ket{v_s}$. We are free to normalise the product $\braket{v_{\ell}|v_{\ell}}=1$, as there is no operator $x \in U(\mathfrak{g})$ such that $x \ket{v_{r}} = \ket{v_{\ell}}$. The vector $\ket{v_s}$ also generates a proper submodule of the module generated by $\ket{v_{\ell}}$, hence the form agrees (and is $0$) on the intersection of the modules generated by $\ket{v_{\ell}}$ and $\ket{v_{r}}$.

It remains to define the logarithmic partner of $\ket{v_s}$ (up to gauge transformations). We choose the action of $L_{0}$ on $\ket{v_h}$ to be
\begin{equation}
    L_{0}\ket{v_{h}} = \ket{v_{s}}.
\end{equation}
The Loewy diagram of ${}^{[0]}P_{0;1,0}$ and a diagram of the gluing action of the algebra are given in Figure~\ref{Fig:M(3,2)1}.
With these choices, the logarithmic couplings which are defined by the relations
\begin{equation}
   G^{-}_{\frac{1}{2}}\ket{v_{h}} = \beta_{\ell}\ket{v_{\ell}}, \quad G^{+}_{\frac{1}{2}}\ket{v_{h}} = \beta_{r} \ket{v_{r}}.
\end{equation}
The values of $\beta_\ell$ and $\beta_r$ follow by determining the action of the algebra on the staggered module ${}^{[0]}P_{0;1,0}$. The quotiented singular vector relations of the head module lift to relations between vectors in the staggered module, which constrain the action of the algebra.

\begin{figure}
\begin{center}
\subfigure
{
\begin{tikzpicture}
\node at (-2.5,2) {${}^{[0]}P_{0;1,0}$};
\node at (2,0) {$L_{1,-\frac{1}{2}}^{\mathrm{NS},-}$};
\node at (0,2) {$L_{0,0}^{\mathrm{NS},+}$};
\node at (-2,0) {$L_{-1,-\frac{1}{2}}^{\mathrm{NS},-}$};
\node at (0,-2) {$L_{0,0}^{\mathrm{NS},+}$};
\draw [thick,->, shorten >= 2mm, shorten <=2mm] (0.4,1.7)--(1.8,0.4);
\draw [thick,->, shorten >= 2mm, shorten <=2mm] (-0.4,1.7)--(-1.8,0.4);
\draw [thick,<-, shorten >= 2mm, shorten <=2mm] (-0.4,-1.7)--(-1.8,-0.4);
\draw [thick,<-, shorten >= 2mm, shorten <=2mm] (0.4,-1.7)--(1.8,-0.4);
\end{tikzpicture}
}
\subfigure
{
\begin{tikzpicture}
\node at (0,0) (H) {$\ket{v_{h}}$};
\node at (0,-1) (S) {$\ket{v_{s}}$};
\node at (2,1) (R) {$\ket{v_{r}}$};
\node at (-2,1) (L) {$\ket{v_{\ell}}$};
\node at (-2,1.5) {$\left(-1,-\tfrac{1}{2}\right)$};
\node at (2.2,1.5) {$\left(1,-\tfrac{1}{2}\right)$};
\node at (0,-1.5) {$(0,0)$};
\draw[thick,->] (H)--(S);
\draw[->] (H) to [out = 70, in=180] (R);
\draw[->] (R) to [out = 270, in=0] (S);
\draw[->] (H) to [out = 110, in=0] (L);
\draw[->] (L) to [out = 270, in=180] (S);
\node[right] at (0,-0.5) {$L_{0}$};
\node[below right] at (0.5,1) {$G^{+}_{\frac{1}{2}}$};
\node[below left] at (-0.5,1) {$G^{-}_{\frac{1}{2}}$};
\node[left] at (-1,-1) {$G^{+}_{-\frac{1}{2}}$};
\node[right] at (1,-1) {$\ G^{-}_{-\frac{1}{2}}$};
\end{tikzpicture}
}
\end{center}
\caption{\textit{Loewy diagram and weight-space diagram of the module ${}^{[0]}P_{0;1,0}$. Both $\ket{v_{h}}$ and $\ket{v_{s}}$ have $L_{0}$ eigenvalue $0$, and $J_{0}$ eigenvalue $0$. Included in the weight-space diagram is the action of the algebra, with the chosen normalisation.}}
\label{Fig:M(3,2)1}
\end{figure}
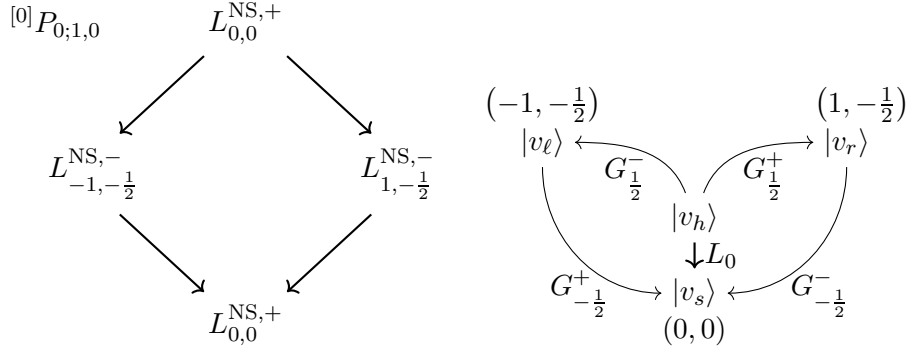

The vectors $G^{\pm}_{-\frac{1}{2}}\ket{0;0}^+$ are singular in the Verma module $V^{\mathrm{NS},+}_{0,0}$, and have been set to $0$ in the irreducible component modules $L^{\NS,+}_{0,0}$. In the staggered module, this implies that
\begin{equation}
    G^{\pm}_{-\frac{1}{2}}\ket{v_{s}} = 0,
\end{equation}
as the socle component module is the submodule isomorphic to $L^{\NS,+}_{0,0}$. Furthermore, the singular vectors of the irreducible component module corresponding to the head give rise to the following relations in the staggered module:
\begin{equation}
    G^{+}_{-\frac{1}{2}}\ket{v_{h}} = - \left( \alpha_{1} L_{-1} + \alpha_{2} J_{-1} \right)\ket{v_{r}}, \qquad
    G^{-}_{-\frac{1}{2}}\ket{v_{h}} = - \left( \gamma_{1} L_{-1} + \gamma_{2} J_{-1} \right)\ket{v_{\ell}},
\end{equation}
for some $\alpha_{1}, \alpha_{2}, \gamma_{1}, \gamma_{2} \in \C$, to be determined. Correspondingly, we introduce vectors of the staggered module as follows:
\begin{equation}
    \ket{\chit^{-}} =  G^{-}_{-\frac{1}{2}}\ket{v_{h}} + \left( \gamma_{1} L_{-1} + \gamma_{2} J_{-1} \right)\ket{v_{\ell}}, \quad \ket{\chit^{+}} = G^{+}_{-\frac{1}{2}}\ket{v_{h}} + \left( \alpha_{1} L_{-1} + \alpha_{2} J_{-1} \right)\ket{v_{r}}.
\end{equation}
Relations resulting from the vanishing of these vectors then constrain the action of the algebra on ${}^{[0]}P_{0;1,0}$.

To illustrate, acting with $L_{1}$ on $\ket{\chit^\pm}$, we have
\begin{equation}
     L_{1}\ket{\chit^{-}} = L_{1} G^{-}_{-\frac{1}{2}}\ket{v_{h}} + L_{1}\left( \gamma_{1} L_{-1} + \gamma_{2} J_{-1} \right)\ket{v_{\ell}}
    = (\beta_{\ell} - \gamma_{1} - \gamma_{2})\ket{v_{\ell}}
\end{equation}
and
\begin{equation}
L_{1}\ket{\chit^{+}}= L_{1} G^{+}_{-\frac{1}{2}}\ket{v_{h}} + L_{1}\left( \alpha_{1} L_{-1} + \alpha_{2} J_{-1} \right)\ket{v_{r}}
    = (\beta_{r}- \alpha_{1} + \alpha_{2})\ket{v_{r}},
\end{equation}
which leads to the relations
\begin{equation}
\beta_{\ell} - \gamma_{1} - \gamma_{2} = 0, \qquad \beta_{r} - \alpha_{1} + \alpha_{2} = 0.
\end{equation}
Similarly, acting with $J_{1}$ leads to
\begin{equation}
    -\beta_{\ell} - \gamma_{1} - \tfrac{1}{3}\gamma_{2}  = 0, \qquad  \beta_{r} + \alpha_{1} - \tfrac{1}{3}\alpha_{2} = 0.
\end{equation}
The relation which defines the logarithmic partner vector $L_{0}\ket{v_{h}} = \ket{v_{s}}$ yields additional constraints by observing that
\begin{equation}
    L_{0}\ket{v_{h}} = \tfrac{1}{2}\big( \{ G^{+}_{-\frac{1}{2}}, G^{-}_{\frac{1}{2}} \} + J_{0}\big) \ket{v_{h}} \implies \beta_{\ell} - \alpha_{1} - \alpha_2  = 2,
\end{equation}
and
\begin{equation}
    L_{0}\ket{v_{h}} = \tfrac{1}{2}\big( \{ G^{-}_{-\frac{1}{2}}, G^{+}_{\frac{1}{2}} \} - J_{0}\big) \ket{v_{h}} \implies \beta_{r} - \gamma_{1} + \gamma_{2} = 2.
\end{equation}
Hence, we have an inhomogeneous linear system of equations, the unique solution of which is given by
\begin{equation}
  \alpha_{1} =-\tfrac{2}{3}, \qquad \alpha_{2} = -1, \qquad \gamma_{1} = -\tfrac{2}{3}, \qquad \gamma_{2} = 1
\end{equation}
and
\begin{equation}
\beta_{\ell} = \beta_{r} = \tfrac{1}{3}.
\end{equation}
The logarithmic couplings $\beta_{\ell}$ and $\beta_{r}$ then uniquely determine the action of the minimal-model algebra on the staggered module ${}^{[0]}P_{0;1,0}$, thereby fixing the structure of the module.

\subsection{The module ${}^{[1]}P_{\frac{3}{2};1,1}$}
Our second example is the staggered module ${}^{[1]}P_{\frac{3}{2};1,1}$, which is a module over the Ramond algebra.
The Loewy diagram and weight-space diagram of the gluing action are given in Figure~\ref{Fig:M(3,2)2}.

We begin with $\ket{v_{r}}$, a fermionic
vector of weight $\left(\tfrac{3}{2},-\tfrac{3}{8}\right)$, which we normalise such that $\braket{v_{r}|v_{r}} = 1$. The vector
\begin{equation}
    \big( \tfrac{1}{3}G^{-}_{-1} - J_{-1}G^{-}_{0}\big) \ket{\tfrac{3}{2},-\tfrac{3}{8}}^-
\end{equation}
is subsingular in the Verma module $V_{\frac{3}{2},-\frac{3}{8}}^{\mathrm{R},-}$, becoming singular after quotienting by the submodule generated by the singular vector
\begin{equation}
    \big( G^{+}_{-1}G^{-}_{0} + J_{-1} + \tfrac{2}{3}L_{-1}\big) \ket{\tfrac{3}{2},-\tfrac{3}{8}}^-.
\end{equation}
We therefore set
\begin{equation}
    \ket{v_{s}} = U_r \ket{v_{r}}, \qquad U_r = \tfrac{1}{3}G^{-}_{-1} - J_{-1}G^{-}_{0}.
\end{equation}
We moreover define $\ket{v_{\ell}}$ by setting $G^{+}_{-1}\ket{v_{\ell}} = \ket{v_{s}}$ and normalising it so that $\braket{v_{\ell} | v_{\ell}} = 1$.

Finally, the logarithmic partner of $\ket{v_{s}}$ must satisfy
\begin{equation}
    L_{0}\ket{v_{h}} = \tfrac{5}{8}\ket{v_{h}} + \ket{v_{s}}.
\end{equation}
The logarithmic couplings are then defined by the relations
\begin{equation}
    \big( \tfrac{1}{3}G^{+}_{1} - G^{+}_{0}J_{1}\big)\ket{v_{h}} = \beta_{r} \ket{v_{r}}, \qquad G^{-}_{1}\ket{v_{h}} = \beta_{\ell}\ket{v_{\ell}}.
\end{equation}

\begin{figure}
\begin{center}
\subfigure
{
\begin{tikzpicture}
\node at (-2.5,2) {${}^{[1]}P_{\frac{3}{2};1,1}$};
\node at (2,0) {$L_{\frac{3}{2},-\frac{3}{8}}^{\mathrm{R},-}$};
\node at (0,2) {$L_{\frac{1}{2},\frac{5}{8}}^{\mathrm{R},+}$};
\node at (-2,0) {$L_{-\frac{1}{2},-\frac{3}{8}}^{\mathrm{R},-}$};
\node at (0,-2) {$L_{\frac{1}{2},\frac{5}{8}}^{\mathrm{R},+}$};
\draw [thick,->, shorten >= 2mm, shorten <=2mm] (0.4,1.7)--(1.8,0.4);
\draw [thick,->, shorten >= 2mm, shorten <=2mm] (-0.4,1.7)--(-1.8,0.4);
\draw [thick,<-, shorten >= 2mm, shorten <=2mm] (-0.4,-1.7)--(-1.8,-0.4);
\draw [thick,<-, shorten >= 2mm, shorten <=2mm] (0.4,-1.7)--(1.8,-0.4);
\end{tikzpicture}
}
\subfigure
{
\begin{tikzpicture}
\node at (0,0) (H) {$\ket{v_{h}}$};
\node at (0,-1) (S) {$\ket{v_{s}}$};
\node at (2,1) (R) {$\ket{v_{r}}$};
\node at (-2,1) (L) {$\ket{v_{\ell}}$};
\node at (-2,1.5) {$\left(-\frac{1}{2},-\frac{3}{8}\right)$};
\node at (2.2,1.5) {$\left(\frac{3}{2},-\frac{3}{8}\right)$};
\node at (0,-1.5) {$\left(\frac{1}{2},\frac{5}{8}\right)$};
\draw[thick,->] (H)--(S);
\draw[->] (H) to [out = 70, in=180] (R);
\draw[->] (R) to [out = 270, in=0] (S);
\draw[->] (H) to [out = 110, in=0] (L);
\draw[->] (L) to [out = 270, in=180] (S);
\node[right] at (0,-0.5) {$L_{0}$};
\node[below right] at (0.5,1) {$U_{r}^{\dagger}$};
\node[below left] at (-0.5,1) {$G^{-}_{1}$};
\node[left] at (-1,-1) {$G^{+}_{-1}$};
\node[right] at (1,-1) {$U_{r}$};
\end{tikzpicture}
}
\end{center}
\caption{\textit{Loewy diagram and weight-space diagram for the module ${}^{[1]}P_{\frac{3}{2};1,1}$. Here, $U_r = \frac{1}{3} G^-_{-1} - J_{-1} G^-_0$.}}
\label{Fig:M(3,2)2}
\end{figure}
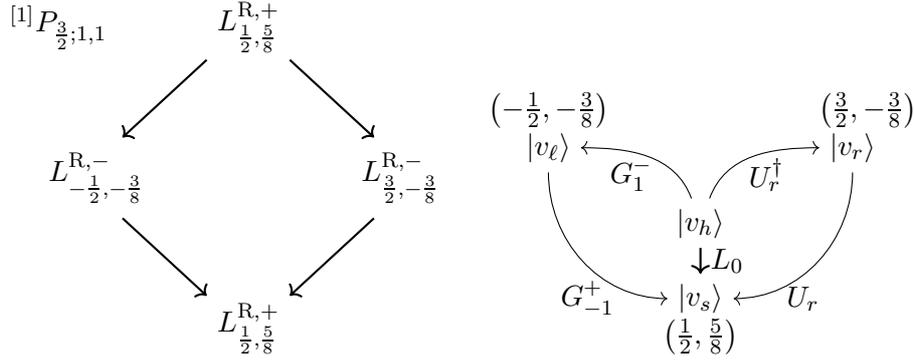

Solving for the logarithmic couplings proceeds as in the previous example. The Verma module $V_{\frac{1}{2},\frac{5}{8}}^{\mathrm{R},+}$ has charged singular vectors at level $1$ with relative charges $\pm 1$. These vectors are given by
\begin{equation}
    G^{+}_{-1}\ket{\tfrac{1}{2},\tfrac{5}{8}}, \qquad \big(L_{-1}G^{-}_{0} + \tfrac{1}{2}J_{-1}G^{-}_{0} - \tfrac{2}{3}G^{-}_{-1} \big)\ket{\tfrac{1}{2},\tfrac{5}{8}}.
\end{equation}
In the staggered module, we compute that the following vectors are zero:
\begin{equation}
	\begin{gathered}
		\ket{\chit^{-}} = \big(L_{-1}G^{-}_{0} + \tfrac{1}{2}J_{-1}G^{-}_{0} - \tfrac{2}{3}G^{-}_{-1} \big) \ket{v_{h}} - \tfrac{32}{101} \big( 2 J_{-2} - \tfrac{8}{9} L_{-2} + \tfrac{1}{3}J_{-1}^{2} \big)\ket{v_{\ell}}, \\
	  \ket{\chit^{+}} =  G^{+}_{-1}\ket{v_{h}} -
		\tfrac{2}{1515}\big( 3819 G^{-}_{-1}G^{+}_{-1} + 1476 J_{-2} - 6982 L_{-2} + 1599 G^{+}_{-2}G^{-}_{0} + 246 J_{-1}^2 \big)\ket{v_{r}}.
	\end{gathered}
\end{equation}
Using these relations, we determine that the generators $\{ G^{-}_{1}, G^{+}_{0}, J_{1} \}$, which generate the positive-mode subalgebra, act on $\ket{v_{h}}$ as
\begin{align}
    G^{+}_{0}\ket{v_{h}} = -\tfrac{542}{101}\big(2L_{-1} + J_{-1} \big)\ket{v_{r}}, \qquad J_{1} \ket{v_{h}} = \tfrac{624}{101} G^{-}_{0} \ket{v_{r}}, \qquad G^{-}_{1}\ket{v_{h}} = \tfrac{32}{303}\ket{v_{\ell}}.
\end{align}

We also have
\begin{equation}
    U^{\dagger}_{r}\ket{v_{h}} =\big( \tfrac{1}{3}G^{+}_{1} - G^{+}_{0}J_{1}\big)\ket{v_{h}}
    = -\tfrac{542}{303}J_{1}\big(J_{-1} + 2L_{-1} \big)\ket{v_{r}} - \tfrac{2496}{303} G^{+}_{0}G^{-}_{0} \ket{v_{r}}
    = \tfrac{656}{909}\ket{v_{r}}.
\end{equation}

Thus, the values of the logarithmic couplings for ${}^{[1]}P_{\frac{3}{2};1,1}$ are
\begin{equation}
    \beta_{\ell} = \frac{32}{303}, \qquad \beta_{r} = \frac{656}{909}.
\end{equation}

\section{Symmetries of $N=2$ staggered module families}
\label{Sec:GenSyms}
The examples above present the first concrete analyses of staggered modules over $N=2$ superconformal algebras. However, the coset provides an infinite number of staggered modules ${}^{[i]}P_{p;r,s}$, labelled by $r,s,i,p$, for each non-unitary minimal model. To better understand the full spectrum of staggered modules, this section investigates their symmetries.

First, we discuss spectral flows of staggered modules and describe the values of $r,s,i,p$ for which two staggered modules are related by spectral flow. Along with this, we analyse the action of spectral flow on the value of the logarithmic couplings describing the module structure. Finally, we identify relations between Loewy diagrams of staggered modules for particular values of the $r,s$ labels.

\subsection{Spectral flow}
\label{Subsec:flow}
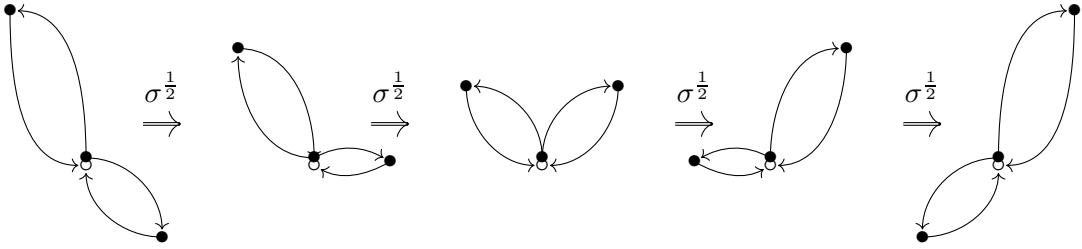
\begin{figure}
   \begin{center}
    \begin{tikzpicture}
\node[] at (0,-1.05) {$\circ$};
\node[] at (0,-0.95) {$\bullet$};
\node[] at (1,0) {$\bullet$};
\node[] at (-1,0) {$\bullet$};
\draw[shorten <=1mm,<-] (1,0) to [out=180,in=90] (0,-0.95);
\draw[shorten <=1mm,<-] (-1,0) to [out=0,in=90] (0,-0.95);
\draw[shorten >=1mm,->] (1,0) to [out=270,in=0] (0,-1.05);
\draw[shorten >=1mm,->] (-1,0) to [out=270,in=180] (0,-1.05);

\node[] at (3,-1.05) {$\circ$};
\node[] at (3,-0.95) {$\bullet$};
\node[] at (4,0.5) {$\bullet$};
\node[] at (2,-1) {$\bullet$};
\draw[shorten <=1mm,<-] (4,0.5) to [out=180,in=90] (3,-0.95);
\draw[shorten <=1mm,<-] (2,-1) to [out=30,in=150] (3,-0.95);
\draw[shorten >=1mm,->] (4,0.5) to [out=270,in=0] (3,-1.05);
\draw[shorten >=1mm,->] (2,-1) to [out=330,in=210] (3,-1.05);

\node[] at (-3,-1.05) {$\circ$};
\node[] at (-3,-0.95) {$\bullet$};
\node[] at (-4,0.5) {$\bullet$};
\node[] at (-2,-1) {$\bullet$};
\draw[shorten <=1mm,<-] (-4,0.5) to [out=270,in=180] (-3,-0.95);
\draw[shorten <=1mm,<-] (-2,-1) to [out=150,in=30] (-3,-0.95);
\draw[shorten >=1mm,->] (-4,0.5) to [out=0,in=90] (-3,-1.05);
\draw[shorten >=1mm,->] (-2,-1) to [out=210,in=330] (-3,-1.05);

\node[] at (6,-1.05) {$\circ$};
\node[] at (6,-0.95) {$\bullet$};
\node[] at (7,1) {$\bullet$};
\node[] at (5,-2) {$\bullet$};
\draw[shorten <=1mm,<-] (7,1) to [out=180,in=90] (6,-0.95);
\draw[shorten <=1mm,<-] (5,-2) to [out=90,in=180] (6,-0.95);
\draw[shorten >=1mm,->] (7,1) to [out=270,in=0] (6,-1.05);
\draw[shorten >=1mm,->] (5,-2) to [out=0,in=270] (6,-1.05);

\node[] at (-6,-1.05) {$\circ$};
\node[] at (-6,-0.95) {$\bullet$};
\node[] at (-7,1) {$\bullet$};
\node[] at (-5,-2) {$\bullet$};
\draw[shorten <=1mm,<-] (-7,1) to [out=0,in=90] (-6,-0.95);
\draw[shorten <=1mm,<-] (-5,-2) to [out=90,in=0] (-6,-0.95);
\draw[shorten >=1mm,->] (-7,1) to [out=270,in=180] (-6,-1.05);
\draw[shorten >=1mm,->] (-5,-2) to [out=180,in=270] (-6,-1.05);

\draw[->,double] (-2.25,-0.5) to (-1.75,-0.5);
\node[above] at (-2,-0.35) {$\sigma^{\frac{1}{2}}$};
\draw[->,double] (1.75,-0.5) to (2.25,-0.5);
\node[above] at (2,-0.35) {$\sigma^{\frac{1}{2}}$};
\draw[->,double] (-5.25,-0.5) to (-4.75,-0.5);
\node[above] at (-5,-0.35) {$\sigma^{\frac{1}{2}}$};
\draw[->,double] (4.75,-0.5) to (5.25,-0.5);
\node[above] at (5,-0.35) {$\sigma^{\frac{1}{2}}$};

\end{tikzpicture}
    \end{center}
    \caption{ \textit{The action of twisting by spectral flow on the relative weights of the component modules. The highest-weight vectors of the component modules are represented by $\bullet$, with the exception of that of the socle, which is represented by $\circ$. The arrows demonstrate the action of the algebra gluing the component modules.}}
    \label{Fig:flowgraph}
\end{figure}

We begin by considering the spectral flow of staggered modules. The results of \cite{CKLR16} state that the spectral flow of a staggered module maintains the Loewy diagram; however, the component modules are spectral flows of the initial component modules. The action of spectral flow on the highest-weight vectors of the component modules is presented diagrammatically in Figure \ref{Fig:flowgraph}. By applying \eqref{moduleauts} to ${}^{[i]}P_{p;r,s}$, we determine that the modules $\sigma^{\ell}\left( {}^{[i]}P_{p;r,s} \right)$ and ${}^{[i-2\ell]}P_{p-2\ell;r,s}$ have the same Loewy diagrams. Recall that for ${}^{[i]}P_{p;r,s}$, we have $p \in \lambda_{r,s} + i + 2\Z$ and that $[i]$ is taken modulo $4$. The Loewy diagram for ${}^{[i]}P_{p;r,s}$ is given in Figure~\ref{Fig:GenN=2Stag}. We remark that equivalent Loewy diagrams are not enough to determine an isomorphism of staggered modules.

To each fixed choice of $r,s$, there are infinitely many staggered modules, labelled by $i,p$, obtained by applying spectral flow (and parity reversal).
We will refer to the set of modules related by spectral flow, for a choice of $r,s$, as a \textit{family}.

We would like to understand the effect of spectral flow on the logarithmic couplings. Consider the relations \eqref{betarels} under the action of the spectral flow. Without loss of generality, we consider the right module only, obtaining
\begin{align}
  \sigma^{\ell}(U_r) \sigma^{\ell}(\ket{v_h}) = \sigma^{\ell}\left( U_{r} \ket{v_{h}} \right) = \sigma^{\ell}\left( \beta_{r} \ket{v_{r}} \right) = \beta_{r} \sigma^{\ell}\left( \ket{v_{r}} \right).
\end{align}
Unfortunately, it is no longer guaranteed (in fact it can only occur rarely) that $\sigma^{\ell}\left( \ket{v_{h}} \right)$ and $\sigma^{\ell}\left( \ket{v_{r}} \right)$ are the highest-weight vectors of the spectral flow of the head and right component modules, respectively. So the spectrally flowed relation does not describe the gluing of highest-weight vectors of the head- and right-component modules; rather, it describes the coupling between {\em extremal vectors} of the component modules, so named \cite{ST96} as they appear as the outermost vectors of a module in a weight-space diagram.

The extremal vectors of an $N=2$ superconformal Verma module are the highest-weight vector $\ket{v}$ and the states
\begin{equation}
    \ket{x^{-}_n} = G^{-}_{-n-\frac{1}{2}}\cdots G^{-}_{-\frac{3}{2}}G^{-}_{-\frac{1}{2}} \ket{v}, \quad \ket{x^{+}_{n}}  = G^{+}_{-n-\frac{1}{2}}\cdots G^{+}_{-\frac{3}{2}}G^{+}_{-\frac{1}{2}}\ket{v}, \quad n \geq 0,
\end{equation}
for Neveu-Schwarz modules, and
\begin{equation}
    \ket{x^{-}_n} = G^{-}_{-n}\cdots G^{-}_{-1} G^{-}_{0} \ket{v}, \quad \ket{x^{+}_{n}}  =  G^{+}_{-n-1}\cdots G^{+}_{-2}G^{+}_{-1}\ket{v}, \quad n \geq 0,
\end{equation}
for Ramond modules. More formally, these are the states of minimal conformal dimension in each $J_0$-eigenspace. It is clear from the equations above that weight spaces containing an extremal vector are one-dimensional. It may occur that an extremal vector is (sub)singular in the Verma module, in which case, it is set to $0$ in the irreducible module. The corresponding expressions for the extremal vectors of the irreducible module are then a straightforward change of the indices given in the above expressions. Clearly, twisting with spectral flow maps extremal vectors to extremal vectors \cite{ST96}.

\begin{figure}
\begin{center}
    \begin{tikzpicture}
\node[] at (0,0) {$\circ$};
\node[] at (0.5,0.5) {$\circ$};
\node[] at (4.7,-.52) {$\circ$};
\node[] at (5.2,0.48) {$\circ$};
\node[] at (4.5,1) {$\bullet$};
\node[] at (4,0) {$\bullet$};
\node[right] at (0.32,0.1) {$\beta$};
\node[right] at (4.2,0.3) {$\tilde{\beta}$};
\node[right] at (4.9,-.32) {$\beta$};
\node[left] at (-0.5,0) {$P$};
\node[left] at (3.5,0) {$\sigma^{\ell}(P)$};
\draw[dotted,thick,->, shorten <=1mm, shorten >=2mm] (0,0) -- (0.5,0.5);
\draw[dotted,thick,->, shorten <=1mm, shorten >=1mm] (4.7,-.52) -- (5.2,0.48);
\draw[thick,->, shorten <=1mm, shorten >=1mm] (4,0) -- (4.5,1);
\draw[thick,-] (0,0) to [out = 190,in=80] (-1,-2);
\draw[thick,-] (0,0) to [out = 350,in=100] (1,-2);
\draw[thick,-] (0.5,0.5) to [out = 190,in=80] (-0.5,-1.5);
\draw[thick,-] (0.5,0.5) to [out = 350,in=100] (1.5,-1.5);
\draw[thick,-] (4,0) to [out = 190,in=80] (3,-2);
\draw[thick,-] (4,0) to [out = 350,in=100] (5,-2);
\draw[thick,-] (4.5,1) to [out = 190,in=80] (3.5,-1);
\draw[thick,-] (4.5,1) to [out = 350,in=100] (5.5,-1);
\end{tikzpicture}
\caption{\textit{The logarithmic coupling $\beta$ between component highest-weight vectors in the module $P$ is mapped to a numerically identical coupling between extremal vectors in the module $\sigma^{\ell}(P)$. The actual logarithmic coupling $\tilde{\beta}$ of $\sigma^{\ell}(P)$ is determined by raising the relation described by $\beta$ to a relation between component highest-weight vectors in $\sigma^{\ell}(P)$.}}
\label{Fig:betatilde}
\end{center}
\end{figure}
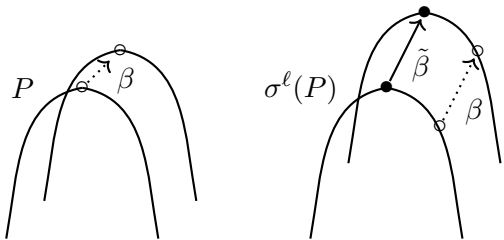

To determine the coupling between highest-weight vectors of component modules after spectral flow, we apply the adjoint raising generator to the coupled extremal vectors. As the action of raising operators on the irreducible component modules is completely determined, the logarithmic couplings of a module ($\beta$) and its spectral flow ($\tilde{\beta}$) are related by a calculable factor. This is presented diagrammatically in Figure~\ref{Fig:betatilde}.

Consider the spectral flow $\sigma^{\ell}(\ket{v})$ of the highest-weight vector $\ket{v}$ of a highest-weight module. Acting with a string of raising generators (without loss of generality, we consider a string of $G^{+}_{r}$ modes), we have
\begin{equation}
    G^{+}_{\frac{1}{2}} G^{+}_{\frac{3}{2}}\cdots G^{+}_{n-\frac{1}{2}} \sigma^{\ell}\left( \ket{v} \right) = \sigma^{\ell}\left(G^{+}_{\frac{1}{2}-\ell}G^{+}_{\frac{3}{2}-\ell} \cdots G^{+}_{n-\ell -\frac{1}{2}}\ket{v} \right).
\end{equation}
If $\ell \geq n$, then the action of the raising operators is generically non-zero and given as above.
If $n > \ell$, then the action
must give zero. When $\ell = n$, we act on an extremal vector with the largest possible string of raising operators, such that the action is non-zero and the resulting vector is extremal. The resulting vector must be proportional to the highest-weight vector of the spectrally-flowed module.

This determines exactly the element of $U(\mathfrak{g})$ one needs to act with to calculate the constant of proportionality for the logarithmic couplings. For spectral flows with $\ell < 0$, the same argument holds with $G^{-}_{r}$ modes, and this carries over straightforwardly to the Ramond sector.

The implication is that it is sufficient to determine the logarithmic couplings for a single module in any given staggered module family (that is, for one choice of $r,s$). As the members of each family are related by spectral flow, the logarithmic couplings of the member modules are all proportional to that of a representative and the constant of proportionality is calculated by commuting the string of raising generators. We shall illustrate this with examples in Section~\ref{Sec:VacuumStag}.
\subsection{Label symmetry}
In the previous section we established that for each pair $r,s$, the staggered modules form a family related by spectral flow. Here, we attempt to determine symmetries of the $r,s$ labels which yield staggered modules with equivalent Loewy diagrams. We refer to these symmetries as \textit{label symmetries}.

It is clear that two component modules ${}^{[i_{1}]}C_{p_{1};r_{1},s_{1}}$ and ${}^{[i_{2}]}C_{p_{2};r_{2},s_{2}}$ can label the same module $L_{j,\Delta}^{\bullet, \pm}$ under the dictionary. We seek to understand for what values of the parameters $p,r,s,i$ the corresponding Loewy diagrams are equivalent up to parity of the component modules.

For representations over the minimal-model algebra $M(3,2)$, there are $4$ possible pairs of $r,s$ labels, namely $(r,s) = (1,0),(1,1),(2,0),(2,1)$. In this setting, we see evidence of general label symmetries. For example,
direct calculation establishes that the modules
${}^{[2]}P_{-2;1,0}$ and ${}^{[0]}P_{-\frac{1}{2};2,1}$, as well as ${}^{[0]}P_{-3;2,0}$ and ${}^{[2]}P_{-\frac{3}{2};1,1}$, have equivalent Loewy diagrams.

Moreover, we note that the modules ${}^{[2]}P_{-2;1,0}$ and  $\sigma^{1}\left({}^{[0]}P_{0;1,0} \right)$, as well as ${}^{[0]}P_{-3;2,0}$ and $\sigma^{2}\left({}^{[0]}P_{1;2,0} \right)$, have equivalent Loewy diagrams. This is suggestive of a symmetry between Loewy diagrams of staggered modules labelled by $(r,0)$ and those labelled by $(u-r,v-1)$.

As each family of staggered modules is related by spectral flow, it suffices to check if the Loewy diagrams of the modules $\sigma^{r} \left({}^{[0]}P_{\lambda_{r,0};r,0}\right)$ and ${}^{[2r+2]}P_{\lambda_{u-r,v-1};u-r,v-1}$ are equivalent, where $\lambda_{r,0} = r-1$ and $\lambda_{u-r,v-1} = -r-1+t$.

\begin{figure}
\begin{center}
\begin{tikzpicture}
\node at (-4,2) {${}^{[-2r]}P_{-r-1;r,0}$};
\node at (1,0) {${}^{[-2r-2]}C_{-r-1-t;r,1}$};
\node at (-1,2) {${}^{[-2r]}C_{-r-1;r,0}$};
\node at (-3,0) {${}^{[-2r]}C_{-r-1+2t;u-r,v-2}$};
\node at (-1,-2) {${}^{[-2r]}C_{-r-1;r,0}$};
\draw [thick,->] (-0.6,1.7)--(0.8,0.4);
\draw [thick,->] (-1.4,1.7)--(-2.8,0.4);
\draw [thick,<-] (-1.4,-1.7)--(-2.8,-0.4);
\draw [thick,<-] (-0.6,-1.7)--(0.8,-0.4);
\node at (3,2) {${}^{[2r+2]}P_{-r-1+t;u-r,v-1}$};
\node at (9,0) {${}^{[2r-2]}C_{-r-1-t;r,1}$};
\node at (7,2) {${}^{[2r+2]}C_{-r-1+t;u-r,v-1}$};
\node at (5,0) {${}^{[2r]}C_{-r-1+2t;u-r,v-2}$};
\node at (7,-2) {${}^{[2r+2]}C_{-r-1+t;u-r,v-1}$};
\draw [thick,->] (7.4,1.7)--(8.8,0.4);
\draw [thick,->] (6.6,1.7)--(5.2,0.4);
\draw [thick,<-] (6.6,-1.7)--(5.2,-0.4);
\draw [thick,<-] (7.4,-1.7)--(8.8,-0.4);
\end{tikzpicture}
\caption{\textit{A comparison of Loewy diagrams for the staggered modules $\sigma^{r}\left({}^{[0]}P_{\lambda_{r,0};r,0}\right) \cong {}^{[-2r]}P_{-r-1;r,0} $ and ${}^{[2r+2]}P_{-r-1+t;u-r,v-1}$. We have used the identifications ${}^{[-2r+2]}C_{-r-1+t;r,-1}\cong {}^{[-2r]}C_{-r-1+2t;u-r,v-2}$ and ${}^{[2r]}C_{-r-1;u-r,v}\cong {}^{[2r-2]}C_{-r-1-t;r,1}$ to simplify the component modules so that the module dictionary can be applied.}}
\label{Fig:r0}
\end{center}
\end{figure}
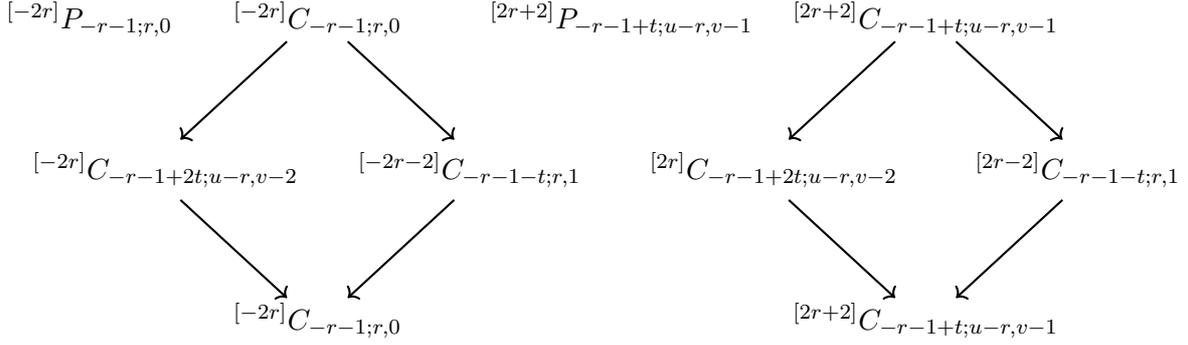

We begin by identifying the Loewy diagrams of the modules $\sigma^{r} \left({}^{[0]}P_{r-1;r,0}\right)$ and ${}^{[-2r]}P_{-r-1;r,0}$, using \eqref{moduleauts}. We can see directly from Figure~\ref{Fig:r0} that the left- and right-component modules of ${}^{[-2r]}P_{-r-1;r,0}$ and ${}^{[2r+2]}P_{-r-1+t;u-r,v-1}$ are isomorphic. It remains to be shown that the head and socle modules are also isomorphic, that is
\begin{equation}
    {}^{[-2r]}C_{-r-1;r,0} \cong {}^{[2r+2]}C_{-r-1+t;u-r,v-1}.
    \label{headiso}
\end{equation}

Applying the module dictionary to ${}^{[-2r]}C_{-r-1;r,0}$, the weights of the corresponding module are given by
\begin{equation}
    j = \frac{-r-1}{t}+1, \qquad \Delta = \Delta^{N=2}_{-r-1;r,0} + \frac{1}{2} = \frac{-r-1}{2t} + \frac{1}{2}.
\end{equation}
Similarly, applying the module dictionary to ${}^{[2r+2]}C_{-r-1+t;u-r,v-1}$, we have that the corresponding weights are given by
\begin{equation}
    j = \frac{-r-1+t}{t}, \qquad \Delta = \Delta^{N=2}_{-r-1+t;u-r,v-1} = \frac{-r-1}{2t} + \frac{1}{2}.
\end{equation}
The weights of the modules are indeed equal, thereby establishing the isomorphism \eqref{headiso} and hence, an equivalence of the Loewy diagrams of the corresponding staggered module families. Moreover, we note that the parities of the component modules are also equal.

We remark that label symmetry can also be understood as arising via the coset. The equivalence of Loewy diagrams arises as a consequence of the fact that the spectral flow of an $L$-type $A_{1}(u,v)$-module gives rise to a $D^{\pm}$-type $A_{1}(u,v)$-module.

Although we have established that the Loewy diagrams are equivalent between module families labelled by $(r,0)$ and $(u-r,v-1)$, this does not imply that the corresponding staggered modules are isomorphic. An isomorphism of staggered modules requires that the logarithmic couplings of the modules are also equal. We discuss this point further in Section \ref{Sec:Disc}.

\section{Staggered modules over $M(2,3)$}
\label{Sec:M(2,3)}
Here we consider $N=2$ staggered modules which are coset counterparts to the $A_{1}(2,3)$ staggered modules analysed in \cite{Gab01,CR12}. The minimal-model algebra $M(2,3)$ has $c= -6$. There are three possible labels: $(r,s) = (1,0), (1,1), (1,2)$. We present an example of a module for $(r,s) = (1,0)$ and one for $(1,1)$.

\subsection{The module ${}^{[0]}P_{0;1,0}$}
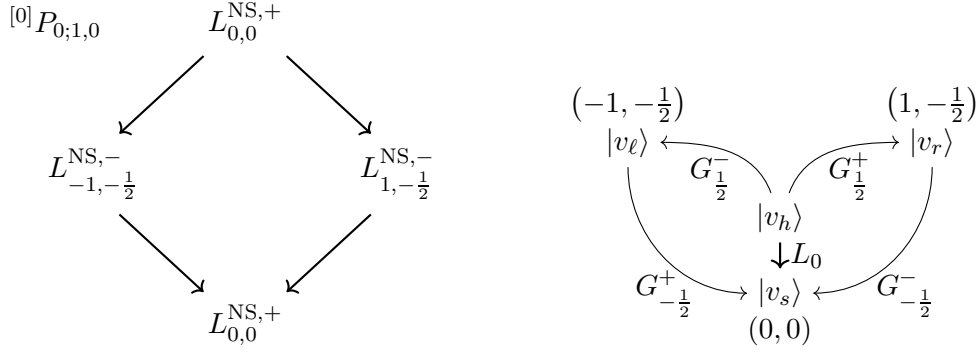
\begin{figure}
\centering
\subfigure
{
\begin{tikzpicture}
\node at (-2.5,2) {${}^{[0]}P_{0;1,0}$};
\node at (2,0) {$L_{1,-\frac{1}{2}}^{\NS,-}$};
\node at (0,2) {$L_{0,0}^{\NS,+}$};
\node at (-2,0) {$L_{-1,-\frac{1}{2}}^{\NS,-}$};
\node at (0,-2) {$L_{0,0}^{\NS,+}$};
\draw [thick,->, shorten >=2mm, shorten <=2mm] (0.4,1.7)--(1.8,0.4);
\draw [thick,->, shorten >=2mm, shorten <=2mm] (-0.4,1.7)--(-1.8,0.4);
\draw [thick,<-, shorten >=2mm, shorten <=2mm] (-0.4,-1.7)--(-1.8,-0.4);
\draw [thick,<-, shorten >=2mm, shorten <=2mm] (0.4,-1.7)--(1.8,-0.4);
\end{tikzpicture}
}
\hspace{1cm}
\subfigure
{
\begin{tikzpicture}
\node at (0,0) (H) {$\ket{v_{h}}$};
\node at (0,-1) (S) {$\ket{v_{s}}$};
\node at (2,1) (R) {$\ket{v_{r}}$};
\node at (-2,1) (L) {$\ket{v_{\ell}}$};
\node at (-2,1.5) {$\left(-1,-\tfrac{1}{2}\right)$};
\node at (2,1.5) {$\left(1,-\tfrac{1}{2}\right)$};
\node at (0,-1.5) {$(0,0)$};
\draw[thick,->] (H)--(S);
\draw[->] (H) to [out = 70, in=180] (R);
\draw[->] (R) to [out = 270, in=0] (S);
\draw[->] (H) to [out = 110, in=0] (L);
\draw[->] (L) to [out = 270, in=180] (S);
\node[right] at (0,-0.5) {$L_{0}$};
\node[below right] at (0.5,1) {$G^{+}_{\frac{1}{2}}$};
\node[below left] at (-0.5,1) {$G^{-}_{\frac{1}{2}}$};
\node[left] at (-1,-1) {$G^{+}_{-\frac{1}{2}}$};
\node[right] at (1,-1) {$\ G^{-}_{-\frac{1}{2}}$};
\end{tikzpicture}
}
\caption{\textit{The Loewy diagram and weight-space diagram for the module ${}^{[0]}P_{0;1,0}$.}}
\label{Fig:10M23}
\end{figure}

We begin with the module ${}^{[0]}P_{0;1,0}$ over the Neveu-Schwarz algebra. We choose vectors according to the same procedure as in the examples in Section~\ref{Sec:M(3,2)}, and present the definitions collectively to shorten the exposition:
\begin{align}
    L_{0}\ket{v_{h}} = \ket{v_{s}}, \quad G^{-}_{\frac{1}{2}}\ket{v_{h}} = \beta_{\ell}\ket{v_{\ell}}, \quad G^{+}_{\frac{1}{2}}\ket{v_{h}} = \beta_{r} \ket{v_{r}}, \quad G^{+}_{-\frac{1}{2}}\ket{v_{\ell}} = \ket{v_{s}}, \quad G^{-}_{-\frac{1}{2}}\ket{v_{r}} = \ket{v_{s}}.
\end{align}
The Loewy diagram and weight-space diagram are given in Figure~\ref{Fig:10M23}.

The quotiented singular vectors in the head component module are $G^{\pm}_{-\frac{1}{2}}\ket{v_{h}}$. Solving as in Section~\ref{Sec:M(3,2)}, the relations in the staggered module are
\begin{equation}
    G^{-}_{-\frac{1}{2}}\ket{v_{\ell}} = \big(\tfrac{3}{2}L_{-1} - J_{-1} \big)\ket{v_{\ell}}, \qquad G^{+}_{-\frac{1}{2}}\ket{v_{h}} = \big(\tfrac{3}{2}L_{-1} + J_{-1} \big) \ket{v_{r}}.
\end{equation}
Using these relations, in conjunction with those coming from
\begin{equation}
    \{ G^{+}_{\frac{1}{2}}, G^{-}_{-\frac{1}{2}} \} \ket{v_{h}} = \left( 2L_{0} + J_{0}\right) \ket{v_{h}}, \quad \{ G^{+}_{-\frac{1}{2}}, G^{-}_{\frac{1}{2}} \} \ket{v_{h}} = \left( 2L_{0} - J_{0}\right) \ket{v_{h}},
\end{equation}
we determine that
\begin{equation}
    G^{-}_{\frac{1}{2}}\ket{v_{h}} = -\tfrac{1}{2}\ket{v_{\ell}}, \qquad G^{+}_{\frac{1}{2}}\ket{v_{h}} =-\tfrac{1}{2} \ket{v_{r}},
\end{equation}
and hence
\begin{equation}
    \beta_{\ell} = \beta_{r} = -\frac{1}{2}.
\end{equation}

\subsection{The module ${}^{[1]}P_{\frac{1}{3};1,1}$}
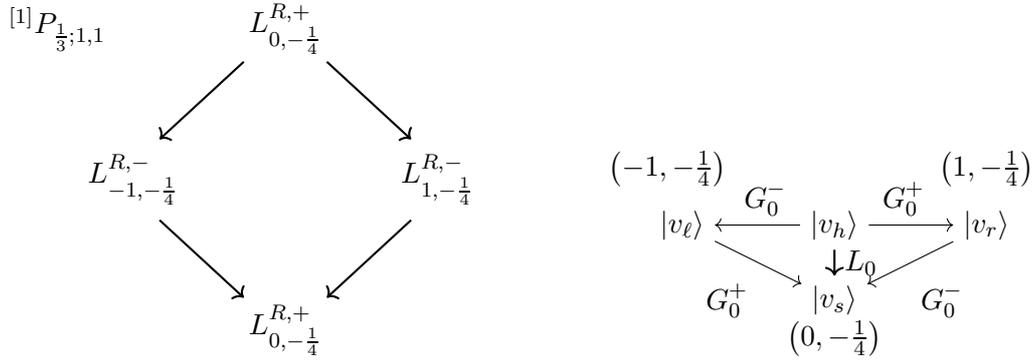
\begin{figure}
\begin{center}
\subfigure{
\begin{tikzpicture}
\node at (-3,2) {${}^{[1]}P_{\frac{1}{3};1,1}$};
\node at (2,0) {$L_{1,-\frac{1}{4}}^{R,-}$};
\node at (0,2) {$L_{0,-\frac{1}{4}}^{R,+}$};
\node at (-2,0) {$L_{-1,-\frac{1}{4}}^{R,-}$};
\node at (0,-2) {$L_{0,-\frac{1}{4}}^{R,+}$};
\draw [thick,->, shorten >=2mm, shorten <=2mm] (0.4,1.7)--(1.8,0.4);
\draw [thick,->, shorten >=2mm, shorten <=2mm] (-0.4,1.7)--(-1.8,0.4);
\draw [thick,<-, shorten >=2mm, shorten <=2mm] (-0.4,-1.7)--(-1.8,-0.4);
\draw [thick,<-, shorten >=2mm, shorten <=2mm] (0.4,-1.7)--(1.8,-0.4);
\end{tikzpicture}
}
\hspace{1cm}
\subfigure
{
\begin{tikzpicture}
\node at (0,0) (H) {$\ket{v_{h}}$};
\node at (0,-1) (S) {$\ket{v_{s}}$};
\node at (2,0) (R) {$\ket{v_{r}}$};
\node at (-2,0) (L) {$\ket{v_{\ell}}$};
\node at (-2.2,0.75) {$\left(-1,-\tfrac{1}{4}\right)$};
\node at (2,0.75) {$\left(1,-\tfrac{1}{4}\right)$};
\node at (0,-1.5) {$\left(0,-\tfrac{1}{4}\right)$};
\draw[thick,->] (H)--(S);
\draw[->] (H) to [out = 0, in = 180] (R);
\draw[->] (R) to (S);
\draw[->] (H) to [out = 180, in=0] (L);
\draw[->] (L) to (S);
\node[right] at (0,-0.5) {$L_{0}$};
\node[below right] at (0.5,0.7) {$G^{+}_{0}$};
\node[below left] at (-0.5,0.7) {$G^{-}_{0}$};
\node[left] at (-1,-1) {$G^{+}_{0}$};
\node[right] at (1,-1) {$G^{-}_{0}$};
\end{tikzpicture}
}
\caption{\textit{The Loewy diagram and weight-space diagram for the module ${}^{[1]}P_{\frac{1}{3};1,1}$.}}
\label{Fig:11M23}
\end{center}
\end{figure}

For the family ${}^{[i]}P_{p;1,1}$, we choose to explore the Ramond sector module ${}^{[1]}P_{\frac{1}{3};1,1}$. The Loewy diagram and weight-space diagram for this module are presented in Figure~\ref{Fig:11M23}.
In this staggered module, all the component modules have $\Delta = \frac{c}{24}$. The vector $G^{-}_{0}\ket{v}$, where $\ket{v}$ is the highest-weight vector, is singular in all Ramond sector Verma modules with $\Delta = \frac{c}{24}$.

For the module ${}^{[1]}P_{\frac{1}{3};1,1}$, we choose the gluing relations to be
\begin{equation}
	\begin{gathered}
	    L_{0}\ket{v_{h}} = -\tfrac{1}{4}\ket{v_{h}} + \ket{v_{s}}, \quad G^{-}_{0}\ket{v_{h}} = \beta_{\ell} \ket{v_{\ell}}, \quad G^{+}_{0} \ket{v_{h}} = \beta_{r}\ket{v_{r}}, \\
	    G^{+}_{0} \ket{v_{\ell}} = \ket{v_{s}}, \quad G^{-}_{0}\ket{v_{r}} = \ket{v_{s}}.
	\end{gathered}
\end{equation}
It is clear from weight-space considerations that
\begin{equation}
    G^{-}_{1}\ket{v_{h}} = J_{1}\ket{v_{h}} = 0.
\end{equation}
As $\Delta = \frac{c}{24}$ for the head module, the charged singular vector $G^{-}_{0}\ket{v_{h}}$ lifts to a relation on the staggered module, namely $G^{-}_{0}\ket{v_{h}} = \beta_{\ell}\ket{v_{\ell}}$. The uncharged quotiented singular vector of the head module appears relatively deep at grade $6$ in the module. As such, determining this vector is computationally challenging.

We find a relation between the logarithmic couplings by considering
\begin{equation}
\{G^{+}_{0},G^{-}_{0} \} \ket{v_{h}} = \left( 2L_{0} + \tfrac{1}{2}\right)\ket{v_{h}}.
\end{equation}
Calculating, we have
\begin{equation}
    \big( 2L_{0} + \tfrac{1}{2} \big) \ket{v_{h}} = 2\ket{v_{s}}
\end{equation}
and
\begin{equation}
    \left(G^{+}_{0}G^{-}_{0} + G^{-}_{0}G^{+}_{0}\right) \ket{v_{h}}
    = G^{+}_{0}\beta_{\ell}\ket{v_{\ell}} + G^{-}_{0}\beta_{r}\ket{v_{r}}
    = \left( \beta_{\ell} + \beta_{r} \right) \ket{v_{s}},
\end{equation}
arriving at
\begin{equation}
    \beta_{\ell} + \beta_{r} = 2.
\end{equation}
The remaining constraint on the logarithmic couplings comes from the observation that the module is conjugation invariant, that is, the Loewy diagrams of $\gamma\left( {}^{[1]}P_{\frac{1}{3};1,1} \right)$ and ${}^{[1]}P_{\frac{1}{3};1,1}$ are equivalent. Using \eqref{gentwisteq} along with the equivalence of Loewy diagrams implies that we may identify $\beta_{\ell} = \beta_{r}$. The action of conjugation simply interchanges their definitions, and maintains the relations which determine their values. Hence, we solve the system of coupled equations and find
\begin{equation}
    \beta_{\ell} = \beta_{r} = 1.
\end{equation}

\section{The module ${}^{[0]}P_{0;1,0}$ over $M(u,v)$}
\label{Sec:VacuumStag}
In Sections~\ref{Sec:M(3,2)} and \ref{Sec:M(2,3)}, we looked at the module ${}^{[0]}P_{0;1,0}$ in the minimal-model algebra with $c = -1$ and $c = -6$, respectively. The weights of the component modules were equal in the two examples. In fact, the component modules have the same weights for all non-unitary $c(t)$ \eqref{admissc}. In this section, we consider the module ${}^{[0]}P_{0;1,0}$ for general $u,v$. As the head component module will be the irreducible highest-weight module $L^{\NS,+}_{0,0}$, we will refer to the staggered module ${}^{[0]}P_{0;1,0}$ as the \textit{vacuum staggered module}.

The head component module of a vacuum staggered module has a quotiented singular vector given by $G^{\pm}_{-\frac{1}{2}}\ket{v_{h}}$. Moreover, the vectors $G^{\pm}_{-\frac{1}{2}}\ket{\mp 1, -\frac{1}{2}}^-$ are singular in $V^{\NS,-}_{\mp1,-\frac{1}{2}}$ for all admissible $c(t)$. Hence, for all vacuum staggered modules, we can choose the gluing action of the algebra to be
\begin{equation}
    L_{0}\ket{v_{h}} = \ket{v_{s}}, \quad G^{-}_{\frac{1}{2}}\ket{v_{h}} = \beta_{r}\ket{v_{\ell}}, \quad G^{+}_{\frac{1}{2}}\ket{v_{h}} = \beta_{r}\ket{v_{r}},\quad G^{+}_{-\frac{1}{2}}\ket{v_{\ell}} = \ket{v_{s}}, \quad G^{-}_{-\frac{1}{2}}\ket{v_{r}} = \ket{v_{s}}.
\end{equation}
As before, the singular vectors of the head module lift to relations in the staggered module:
\begin{equation}
    G^{-}_{-\frac{1}{2}}\ket{v_{h}} = \left( \gamma_{1} L_{-1} + \gamma_{2} J_{-1} \right)\ket{v_{\ell}}, \qquad  G^{+}_{-\frac{1}{2}}\ket{v_{h}} = \left( \alpha_{1} L_{-1} + \alpha_{2} J_{-1} \right)\ket{v_{r}}.
\end{equation}
We compute the action of the same generators as in Section~\ref{Sec:M(3,2)}. For this example, the resulting relations are
dependent on $c$:
\begin{gather}
 \beta_{r} - \alpha_{1} + \alpha_{2}  = 0,  \quad \beta_{r} + \alpha_{1} + \frac{c}{3}\alpha_{2}  = 0, \nn
\beta_{r} - \gamma_{1} - \gamma_{2} = 0, \quad -\beta_{\ell} - \gamma_{1} + \frac{c}{3}\gamma_{2} = 0, \nn
\beta_{\ell} - \alpha_{1} - \alpha_2 = 2, \quad \beta_{r} - \gamma_{1} + \gamma_{2} = 2.
\end{gather}
Solving this system of equations yields
\begin{equation}
    \alpha_{1} = \gamma_{1} = \frac{1}{t}, \quad \alpha_{2} = 1, \quad \gamma_{2} = -1
\end{equation}
and
\begin{equation}
 \beta_{\ell} = \beta_{r} = \frac{t-1}{t}.
\end{equation}

As expected, the logarithmic couplings are indeed functions of $t$. Combining this result with our previous results, we have in principle determined the structure of the staggered module family ${}^{[i]}P_{p,1,0}$ (and equivalently ${}^{[i]}P_{p+t,u-1,v-1}$) over the minimal-model algebras $M(u,v)$.

We can use these modules to better understand the action of spectral flow on the logarithmic couplings. In Section~\ref{Sec:GenSyms}, we noted that the logarithmic couplings change under spectral flow, by factors which can be in principle calculated for any value of the flow parameter $\ell$. The next two examples investigate this directly using the spectral flows of the vacuum staggered modules.

\subsection{The modules $\sigma^{\pm\frac{1}{2}}\big( {}^{[0]}P_{0;1,0}\big)$}
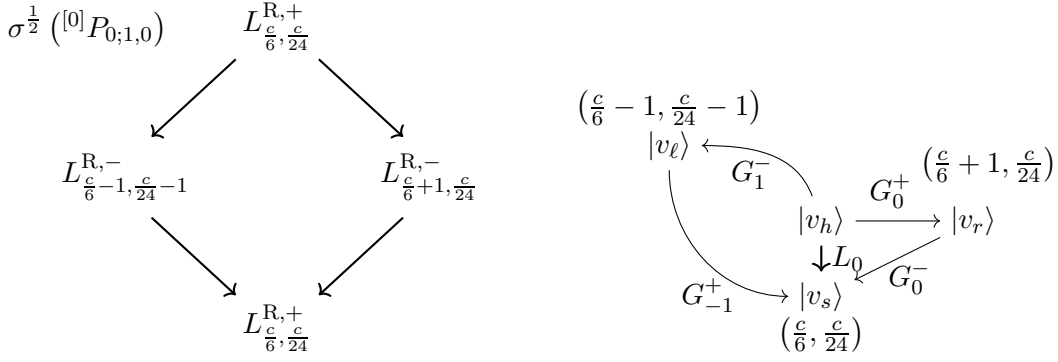
\begin{figure}
\begin{center}
\subfigure
{
\begin{tikzpicture}
\node at (-2.5,2) {$\sigma^{\frac{1}{2}}\left({}^{[0]}P_{0;1,0}\right)$};
\node at (2,0) {$L_{\frac{c}{6}+1,\frac{c}{24}}^{\mathrm{R},-}$};
\node at (0,2) {$L_{\frac{c}{6},\frac{c}{24}}^{\mathrm{R},+}$};
\node at (-2,0) {$L_{\frac{c}{6}-1,\frac{c}{24}-1}^{\mathrm{R},-}$};
\node at (0,-2) {$L_{\frac{c}{6},\frac{c}{24}}^{\mathrm{R},+}$};
\draw [thick,->, shorten >= 2mm, shorten <=2mm] (0.4,1.7)--(1.8,0.4);
\draw [thick,->, shorten >= 2mm, shorten <=2mm] (-0.4,1.7)--(-1.8,0.4);
\draw [thick,<-, shorten >= 2mm, shorten <=2mm] (-0.4,-1.7)--(-1.8,-0.4);
\draw [thick,<-, shorten >= 2mm, shorten <=2mm] (0.4,-1.7)--(1.8,-0.4);
\end{tikzpicture}
}
\hspace{0.5cm}
\subfigure
{
\begin{tikzpicture}
\node at (0,0) (H) {$\ket{v_{h}}$};
\node at (0,-1) (S) {$\ket{v_{s}}$};
\node at (2,0) (R) {$\ket{v_{r}}$};
\node at (-2,1) (L) {$\ket{v_{\ell}}$};
\node at (-2,1.5) {$\big(\tfrac{c}{6}-1,\tfrac{c}{24}-1\big)$};
\node at (2.2,0.75) {$\big(\tfrac{c}{6}+1,\tfrac{c}{24}\big)$};
\node at (0,-1.5) {$\big(\tfrac{c}{6},\tfrac{c}{24}\big)$};
\draw[thick,->] (H)--(S);
\draw[->] (H) to [out = 0, in=180] (R);
\draw[->] (R) to (S);
\draw[->] (H) to [out = 110, in=0] (L);
\draw[->] (L) to [out = 270, in=180] (S);
\node[right] at (0,-0.5) {$L_{0}$};
\node[below right] at (0.5,0.75) {$G^{+}_{0}$};
\node[below left] at (-0.5,1) {$G^{-}_{1}$};
\node[left] at (-1,-1) {$G^{+}_{-1}$};
\node[right] at (0.75,-0.75) {$G^{-}_{0}$};
\end{tikzpicture}
}
\end{center}
\caption{\textit{The Loewy diagram and weight-space diagram for $\sigma^{\frac{1}{2}}\left( {}^{[0]}P_{0;1,0}\right)$.}}
\label{Fig:sig1/2}
\end{figure}

Applying $\sigma^{\frac{1}{2}}$ to the vacuum staggered module ${}^{[0]}P_{0;1,0}$ yields a staggered module over the Ramond algebra, with Loewy and weight-space diagram given in Figure~\ref{Fig:sig1/2}.

For this Ramond-algebra module, we choose the gluing relations and definitions of the logarithmic couplings to be
\begin{gather}
    L_{0}\ket{v_{h}} = \frac{c}{24}\ket{v_{h}} + \ket{v_{s}}, \quad G^{-}_{1}\ket{v_{h}} = \beta_{\ell}\ket{v_{\ell}}, \quad G^{+}_{0}\ket{v_{h}} = \beta_{r}\ket{v_{r}}, \nn
     G^{+}_{-1}\ket{v_{\ell}} = \ket{v_{s}}, \quad G^{-}_{0}\ket{v_{r}} = \ket{v_{s}}.
\end{gather}
Note that the right, head, and socle modules all have $\Delta = \frac{c}{24}$. We remark that this example realises the case where the spectral flow of highest-weight vectors of component modules (those in the vacuum staggered module examples) are mapped to highest-weight vectors of spectral flows of component modules. As noted in Section~\ref{Sec:GenSyms}, spectral flow $\sigma^\ell$ needs only preserve the property of being extremal, but it may happen, for sufficiently small values of $\ell$, that highest-weight vectors are mapped to highest-weight vectors.

The staggered module singular vectors are in this case of the form
\begin{align}
    G^{-}_{0}\ket{v_{h}} = \left( \gamma_{1} L_{-1} + \gamma_{2}J_{-1} \right)\ket{v_{\ell}}, \qquad G^{+}_{-1}\ket{v_{h}} = \left( \alpha_{1} L_{-1} + \alpha_{2}J_{-1} \right)\ket{v_{r}}.
\end{align}
The annihilation conditions of these vectors, along with relations given by
\begin{equation}
    \{G^{+}_{0},G^{-}_{0} \}\ket{v_h}= \left( 2L_{0} - \frac{c}{12}\right)\ket{v_{h}}, \qquad \{G^{+}_{-1},G^{-}_{1} \}\ket{v_h} = \left( 2L_{0} - 2J_{0} + \frac{c}{4} \right) \ket{v_{h}},
\end{equation}
lead to a linear system of equations for the unknown parameters. The resulting set of equations again has a solution for all admissible $c(t)$, given by
\begin{equation}
    \alpha_{1} = \gamma_{1} = \frac{1}{t}, \quad \alpha_{2} = 1 - \frac{1}{2t}, \quad \gamma_{2} = -1 - \frac{1}{2t}
\end{equation}
and
\begin{equation}
 \beta_{\ell} = \beta_{r} = \frac{t-1}{t}.
\end{equation}
We see that indeed $\beta_{\ell}, \beta_{r}$ are unchanged under the action of $\sigma^{\frac{1}{2}}$, as expected from Section~\ref{Subsec:flow}.

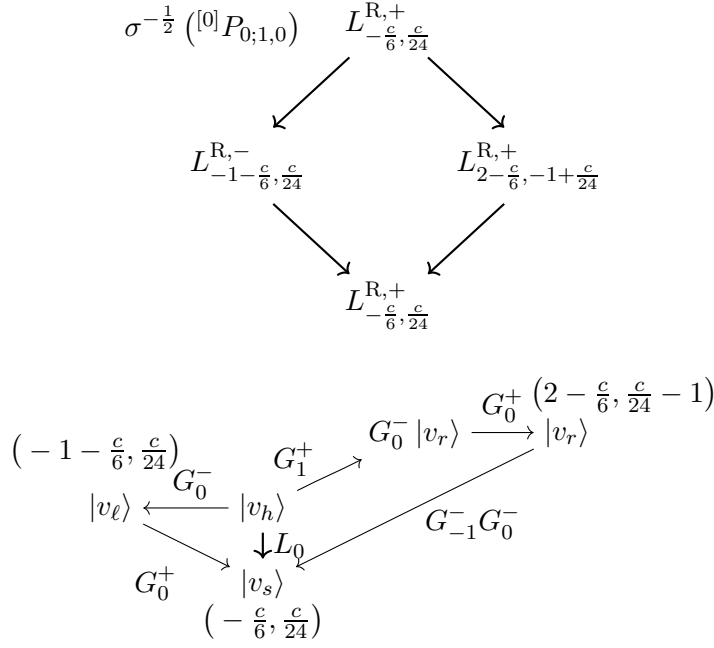
\begin{figure}
\begin{center}
\subfigure{
\begin{tikzpicture}[scale=0.925]
\node at (-2.5,2) {$\sigma^{-\frac{1}{2}}\left( {}^{[0]}P_{0;1,0}\right)$};
\node at (2,0) {$L_{2-\frac{c}{6},-1+\frac{c}{24}}^{\mathrm{R},+}$};
\node at (0,2) {$L_{-\frac{c}{6},\frac{c}{24}}^{\mathrm{R},+}$};
\node at (-2,0) {$L_{-1-\frac{c}{6},\frac{c}{24}}^{\mathrm{R},-}$};
\node at (0,-2) {$L_{-\frac{c}{6},\frac{c}{24}}^{\mathrm{R},+}$};
\draw [thick,->, shorten <= 2mm, shorten >= 2mm] (0.4,1.7)--(1.8,0.4);
\draw [thick,->, shorten <= 2mm, shorten >= 2mm] (-0.4,1.7)--(-1.8,0.4);
\draw [thick,<-,shorten <= 2mm, shorten >= 2mm] (-0.4,-1.7)--(-1.8,-0.4);
\draw [thick,<-,shorten <= 2mm, shorten >= 2mm] (0.4,-1.7)--(1.8,-0.4);
\end{tikzpicture}
}
\hfill
\subfigure
{
\begin{tikzpicture}
\node at (0,0) (H) {$\ket{v_{h}}$};
\node at (0,-1) (S) {$\ket{v_{s}}$};
\node at (2,1) (GR) {$G^{-}_{0}\ket{v_{r}}$};
\node at (4,1) (R) {$\ket{v_{r}}$};
\node at (-2,0) (L) {$\ket{v_{\ell}}$};
\node at (-2.2,0.75) {$\big(-1-\tfrac{c}{6},\tfrac{c}{24}\big)$};
\node at (4.75,1.5) {$\big(2-\tfrac{c}{6},\tfrac{c}{24}-1\big)$};
\node at (0,-1.5) {$\big(-\tfrac{c}{6},\tfrac{c}{24}\big)$};
\draw[thick,->] (H)--(S);
\draw[->] (H) to [out = 27, in=207] (GR);
\draw[->] (GR) to node[above] {$G^+_0$} (R);
\draw[->] (R) to (S);
\draw[->] (H) to [out = 180, in=0] (L);
\draw[->] (L) to (S);
\node[right] at (0,-0.5) {$L_{0}$};
\node[below right] at (0,1) {$G^{+}_{1}$};
\node[below left] at (-0.5,0.7) {$G^{-}_{0}$};
\node[left] at (-1,-1) {$G^{+}_{0}$};
\node[right] at (2,-0.2) {$G^{-}_{-1}G^{-}_{0}$};
\end{tikzpicture}
}
\caption{\textit{The Loewy diagram and weight-space diagram for $\sigma^{-\frac{1}{2}}\left( {}^{[0]}P_{0;1,0}\right)$.}}
\label{Fig:sig-1/2}
\end{center}
\end{figure}

We also consider the module $\sigma^{-\frac{1}{2}}\left({}^{[0]}P_{0;1,0}\right)$ for general admissible $t$. The Loewy diagram and weight-space diagram are given in Figure~\ref{Fig:sig-1/2}. We choose the following gluing:
\begin{gather}
    L_{0}\ket{v_{h}} = \frac{c}{24}\ket{v_{h}} + \ket{v_{s}}, \qquad  G^{-}_{0}\ket{v_{h}} = \beta_{\ell} \ket{v_{\ell}}, \qquad G^{+}_{1}\ket{v_{h}} = \beta_{r}G^{-}_{0}\ket{v_{r}}, \nn
    G^{+}_{0}\ket{v_{\ell}} = \ket{v_{s}}, \qquad G^{-}_{-1}G^{-}_{0}\ket{v_{r}} = \ket{v_{s}}.
\end{gather}
Note, we have set $G^{+}_{1}\ket{v_{h}} = \beta_{r}G^{-}_{0}\ket{v_{r}}$. Applying spectral flow to the relation $G^{+}_{\frac{1}{2}}\ket{v_{h}}_{V} = \beta_{r}\ket{v_{r}}_{V}$ in the vacuum staggered modules (subscript $V$ denoting vacuum staggered module vectors) yields
\begin{equation}
    \sigma^{-\frac{1}{2}}\big(G^{+}_{\frac{1}{2}}\ket{v_{h}}_{V}\big)  = \sigma^{-\frac{1}{2}}\left(\beta_{r}\ket{v_{r}}_{V} \right) = \beta_{r}\sigma^{-\frac{1}{2}}\left(\ket{v_{r}}_{V} \right)  = G^{+}_{1}\sigma^{-\frac{1}{2}}\left(\ket{v_{h}}_{V}\right),
\end{equation}
hence we expect to find $\beta_{r} = \frac{t-1}{t}$. The gluing relation between right and socle component highest-weight vectors is given by $G^{-}_{-1}G^{-}_{0}\ket{v_{r}} = \ket{v_{s}}$. Recalling the notation $\tilde{\beta}_{r}$ of Section~\ref{Subsec:flow} for the logarithmic coupling of the spectrally flowed module (see Figure~\ref{Fig:betatilde}), we have $G^{+}_{0}G^{+}_{1}\ket{v_{h}} = \tilde{\beta}_{r} \ket{v_{r}}$. We identify $G^{+}_{0}$ with the string of raising generators required to map from the image of the highest-weight vector under spectral flow, to the highest-weight vector of the component module.

The relations coming from quotiented singular vectors are
\begin{equation}
    G^{-}_{-1}\ket{v_{h}} = \left( \gamma_{1}L_{-1} + \gamma_{2}J_{-1} \right) \ket{v_{\ell}}, \qquad G^{+}_{0}\ket{v_{h}} = \left(\alpha_{1}L_{-1}G^{-}_{0} + \alpha_{2}J_{-1}G^{-}_{0} \right) \ket{v_{r}}.
\end{equation}
We also require the relations coming from
\begin{equation}
    \{G^{+}_{0},G^{-}_{0} \}\ket{v_{h}} = \left( 2L_{0} - \frac{c}{12}\right)\ket{v_{h}}, \qquad \{G^{+}_{1},G^{-}_{-1} \}\ket{v_{h}} = \left( 2L_{0} + 2J_{0} + \frac{c}{4} \right) \ket{v_{h}}.
\end{equation}
Note that the second anti-commutation relation differs from that in the previous example.

The solutions for the parameter functions are
\begin{equation}
    \alpha_{1} = \gamma_{1} = \frac{1}{t}, \quad \gamma_{2} = -1 + \frac{1}{2t}, \quad \alpha_{2} = 1 + \frac{1}{2t},
\end{equation}
and
\begin{equation}
    \beta_{\ell} = \beta_{r} = \frac{t-1}{t},
\end{equation}
confirming the expected value for $\beta_r$.
To calculate $\tilde{\beta}_{r}$, we use that
\begin{equation}
    G^{+}_{0}G^{+}_{1}\ket{v_{h}} = \beta_{r}G^{+}_{0}G^{-}_{0}\ket{v_{r}} = \frac{t-1}{t}\left(2L_{0} - \frac{c}{12}\right) \ket{v_{r}} = (-2)\frac{t-1}{t}\ket{v_{r}},
\end{equation}
from which it follows that
\begin{equation}
\tilde{\beta}_{r} = -2\frac{t-1}{t}.
\end{equation}
This confirms the general result of Section~\ref{Subsec:flow} that $\tilde{\beta}_{r}$ is related to $\beta_r$, with the proportionality constant determined by the action of the raising operators: in this case, $-2$.

\section{Discussion}
\label{Sec:Disc}
In this paper, we have presented a first investigation of staggered modules over the non-unitary $N=2$ superconformal minimal-model algebras $M(u,v)$. We have provided several concrete examples, along with general results regarding symmetries of the spectral-flow families of staggered modules produced via the coset (\ref{Mcoset}).

A novel feature of the modules investigated here is the gluing of component modules into spaces generated by subsingular vectors, rather than simply those of singular vectors. By applying \eqref{specflowaction} to the weights of the component modules, one sees that this feature occurs generically in staggered $M(u,v)$-modules. The difference between $j$-labels of the head/socle and either left or right irreducible component modules becomes $2$. A difference of $2$ is only possible if the gluing is into subsingular spaces, as all singular vectors for Neveu-Schwarz Verma modules appear with $J_{0}$-weight of $\{ \pm1, 0 \}$ relative to the Verma module highest-weight vector \cite{BFK86}.

We have assumed that the Loewy diagram in conjunction with the logarithmic couplings is sufficient to describe the structure of a given $N=2$ staggered module. This is well-motivated by previous examples of logarithmic CFTs. In particular, the authors of \cite{KR09} proved that the logarithmic couplings of staggered modules over the Virasoro algebra in fact parametrise the space of isomorphism classes of staggered modules. Pursuing a similar result for $N=2$ staggered modules is a natural next step.

The explicit examples of Sections \ref{Sec:M(3,2)}, \ref{Sec:M(2,3)}, and \ref{Sec:VacuumStag}, provide strong evidence that the structure of staggered modules over the $N=2$ minimal-model algebras $M(u,v)$ is uniquely determined by their corresponding Loewy diagrams. If the Loewy diagrams were sufficient to determine the structure, label symmetry would determine an isomorphism of staggered modules (and hence, staggered module families).

This isomorphism of staggered modules would also follow if ${}^{[i]}P_{p;r,s}$ were proven to be the projective cover of the irreducible module ${}^{[i]}C_{p;r,s}$, as is conjectured \cite{CLRW19}. As projective covers are unique, up to isomorphism, equivalence of Loewy diagrams (or simply isomorphisms between head modules) implies the desired isomorphism. Should such an isomorphism exist, then the explicit examples of modules provided in Sections \ref{Sec:M(3,2)} and \ref{Sec:M(2,3)} would be sufficient to understand the structure of all staggered modules over $M(3,2)$ and $M(2,3)$ arising via the coset.

In the paper \cite{CLRW19}, the authors give conjectural fusion rules for the atypical modules over $M(u,v)$ minimal-model algebras, coming from similar results for atypical $A_{1}(u,v)$-modules. With the concrete realisations of staggered modules over the $N=2$ algebras presented here, another natural step would be to check that these staggered modules do indeed arise in these fusion rules for $M(3,2)$ and $M(2,3)$.

The conjectured fusion rules for the $A_{1}(u,v)$ minimal models also involve spectral flows of relaxed highest-weight modules \cite{CR132}. Moreover, staggered modules over $A_{1}(3,2)$ and $A_{1}(2,3)$ have been shown to have conformal weights that are unbounded below. That is, understanding the $A_{1}(u,v)$ minimal models using the standard module formalism requires modules that are not positive-energy.

The coset counterpart $N=2$ modules are necessarily positive-energy as the fermionic generators of the $N=2$ algebra square to zero. As such, it may be easier to determine properties of interest in the $N=2$ setting, for instance genuine fusion rules and whether staggered modules are projective. These results would then lift to similar results for the $A_{1}(u,v)$ minimal models, by the results of \cite{CLRW19,CKLR16,Sato19}.

\section*{Acknowledgements}
CR was supported by a University of Queensland Research Award and by the Australian Research Council under the Discovery Project scheme, numbers DP170103265 and DP200100067.
DR's research is supported by the Australian Research Council Discovery Project DP160101520.
JR was supported by the Australian Research Council under the Discovery Project scheme, project number DP160101376.

\flushleft

\end{document}